\documentclass[prd,superscriptaddress,nofootinbib,notitlepage]{revtex4-1}
\usepackage{epsfig,amsmath,amssymb,slashed}
\usepackage{graphicx}
\usepackage{color}
\usepackage{siunitx}
\usepackage{color}
\usepackage{bm}
\usepackage{subfig}
\newcommand{\bra}[1]{\langle #1|}
\newcommand{\ket}[1]{|#1\rangle}
\newcommand{\braket}[2]{\langle #1|#2\rangle}
\newcommand{\media}[1]{\langle #1 \rangle}
\newcommand{\wad}[1]{\widehat a^\dagger(#1)}
\newcommand{\wa}[1]{\widehat a(#1)}
\newcommand{\wbd}[1]{\widehat b^\dagger(#1)}
\newcommand{\wb}[1]{\widehat b(#1)}
\newcommand{\wadf}[2]{{\widehat a}^\dagger_#1 (#2)}
\newcommand{\waf}[2]{{\widehat a}_#1 (#2)}
\newcommand{\wbdf}[2]{\widehat b^\dagger_{#1}(#2)}
\newcommand{\wbf}[2]{\widehat b_{#1}(#2)}
\newcommand{\di}{{\rm d}}
\newcommand{\ii}{i}

\def\wT{{\widehat T}}
\def\wj{{\widehat j}}
\def\wQ{{\widehat Q}}
\def\wP{{\widehat P}}
\def\wJ{{\widehat J}}
\def\wW{{\widehat W}}
\def\ww{{\widehat w}}
\def\wspt{{\widehat{\cal S}}}

\def\wpsi{{\widehat{\psi}}}
\def\wrho{{\widehat{\rho}}}
\def\wLa{{\widehat{\Lambda}}}
\def\wrhol{{\widehat{\rho}_{\rm LE}}}

\def\wA{{\widehat{A}}}
\def\wB{{\widehat{B}}}

\def\codevmu{{\stackrel{\leftrightarrow}{\partial^\mu}}}
\def\codevnu{{\stackrel{\leftrightarrow}{\partial^\nu}}}
\def\codevi{{\stackrel{\leftrightarrow}{\frac{\partial}{\partial p^i}}}}
\def\codevki{{\stackrel{\leftrightarrow}{\frac{\partial}{\partial k^i}}}}
\newcommand{\tr}{{\rm tr}}  
\newcommand{\Tr}{{\rm Tr}} 
  
\newcommand{\kv}{\textrm{k}}
\newcommand{\e}{{\rm e}}

\newcommand{\p}{{\rm p}}
\newcommand{\x}{{\rm x}}

\newcommand{\Psibar}{{\overline \Psi}}
\newcommand{\be}{\begin{equation}}
\newcommand{\ee}{\end{equation}}                                                                               
\newcommand{\bea}{\begin{eqnarray}}
\newcommand{\eea}{\end{eqnarray}}

\begin{document}

\title{Polarization in relativistic fluids: a quantum field theoretical derivation} 

\author{F. Becattini}
\affiliation{Universit\`a di Firenze and INFN Sezione di Firenze, Florence, Italy}

\begin{abstract}
We review the calculation of polarization in a relativistic fluid within the framework
of statistical quantum field theory. We derive the expressions of the spin density 
matrix and the mean spin vector both for a single quantum relativistic particle 
and for a quantum free field. After introducing the formalism of the covariant Wigner 
function for the scalar and the Dirac field, the relation between spin density matrix 
and the covariant Wigner function is obtained. The formula is applied to the fluid 
produced in relativistic nuclear collisions by using the local thermodynamic 
equilibrium density operator and recovering previously known formulae. The dependence 
of these results on the spin tensor and pseudo-gauge transformations of the stress-energy
tensor is addressed.
\end{abstract}

\maketitle

\section{Introduction}
\label{intro}

The discovery of global polarization of $\Lambda$ hyperons in relativistic nuclear
collisions \cite{STAR:2017ckg,Voloshin:2017kqp,Adam:2018ivw,Niida:2018hfw,Adam:2019srw} 
has sparked a great interest in the theory of spin and 
polarization in relativistic fluids and relativistic matter in general. Several 
approaches have been proposed and several are currently pursued; amongst them, relativistic
kinetic theory \cite{Wang:2019moi,Hattori:2019ahi,Gao:2019znl,Weickgenannt:2019dks,
Yang:2020hri,Liu:2020flb} and a phenomenological treatment of the spin tensor 
\cite{Florkowski:2017ruc,Florkowski:2018fap,Florkowski:2019qdp}.

Yet, the most fundamental tool is quantum statistical field theory, which was used to derive
the original formula \cite{Becattini:2013fla} of polarization of quasi-free particles
in a relativistic fluid at local thermodynamic equilibrium. All other approaches
should, in the first place, reproduce the results obtained with this method, once 
the state of the system, that is its density operator, is chosen.  

Carrying out the calculation of the polarization matrix, which for spin 1/2 particles 
boils down to the mean spin vector, in a quantum field theoretical framework
is, however, not an easy task. In the derivation presented in ref.~\cite{Becattini:2013fla}
some approximation were introduced and the final formula admittedly relied on the 
use of the canonical spin tensor, that is dependent on the specific set
of quantum stress-energy and spin tensor (in other words of a the {\em pseudo-gauge} 
choice \cite{Becattini:2018duy}). It is the purpose of this paper to review the 
derivation of the polarization of spin 1/2 particles step by step and fill some
of the conceptual gaps. Particularly, the exact formula relating the spin density
matrix and the mean spin vector to the covariant Wigner function will be found and
it will be thereby conclusively demonstrated that the {\em expression} of the polarization 
is independent of the spin tensor. Furthermore, a general formula for particles 
with any spin $S$ will be derived in the limit of distinguishable quantum particles, 
that is neglecting quantum statistics. For this purpose, many useful concepts in
statistical quantum field theory will be thoroughly reviewed and discussed.

The paper is organized as follows: in Section \ref{spin} the general definitions
of spin density matrix and mean spin vector will be given in a quantum relativistic 
framework. In Section \ref{single} a formula for the spin density matrix and the
mean spin vector will be derived for a single free quantum particle with spin $S$
in a thermal bath with rotation and acceleration, by using only group theory techniques.
In Section \ref{wigner} the covariant Wigner operator and function will be introduced
for the free scalar and Dirac field. In Section \ref{polwig} the 
formula for the mean spin vector of a free Dirac fermion will be obtained as a function
of the covariant Wigner function while in Section \ref{polang} the same formula
will be obtained with a different method based on total angular momentum, already
used in ref.~\cite{Becattini:2013fla}. In Section \ref{wigloc} the density operator 
at local thermodynamic equilibrium will be discussed in detail
with emphasis on its application in relativistic heavy ion collisions and the 
derivation of the formula of mean spin vector for a fermion in a relativistic fluid 
at local thermodynamic equilibrium will be outlined.

\subsection*{Notations and conventions}

In this paper we use the natural units, with $\hbar=c=K=1$.\\
We will use the relativistic notation with repeated indices assumed to be saturated, 
however, contractions of indices will be sometimes denoted with, e.g. $\beta_\mu p^\mu
= \beta \cdot p$. The Minkowskian metric tensor is ${\rm diag}(1,-1,-1,-1)$; for 
the Levi-Civita symbol we use the convention $\epsilon^{0123}=1$.\\  
Operators in Hilbert space will be denoted by a large upper hat ($\wT$) while 
unit vectors with a small upper hat ($\hat v$). Noteworthy exeception, the
Dirac field which is expressed by $\Psi$ without an upper hat. \\
The symbol $\Tr$ with a capital T stands for the trace over all states in 
the Hilbert space, whereas the symbol $\tr$ stands for a trace over polarization
states or traces of finite dimensional matrices.

\section{The spin density matrix and the definition of mean spin}
\label{spin}

In relativistic quantum mechanics, for a single massive particle, the spin angular 
momentum vector is defined as:
\begin{equation}\label{luba}
 \widehat S^\mu = -\frac{1}{2m} \epsilon^{\mu \nu \rho \sigma} \wJ_{\nu \rho} 
 \widehat P_\sigma 
\end{equation} 
where $\wJ_{\nu \rho}$ are the angular momentum-boost operators and $\wP_\sigma$ 
the energy-momentum operator. The operator in eq.~\eqref{luba} is also known as 
Pauli-Lubanski vector and it fulfills the following commutation relations:
\begin{eqnarray}
&& [\widehat S_\mu, \widehat P_\nu] = 0 \\ \nonumber
&& [\widehat S_\mu, \widehat S_\nu] = -i \epsilon_{\mu \nu \rho \sigma} 
   \widehat S^\rho \widehat P^\sigma \\ \nonumber
&& \widehat S \cdot \widehat P = 0
\end{eqnarray}
Hence, if the ket $\ket{p}$ is an eigenvector of $\widehat P$, so is $\widehat S \ket{p}$. 
The restriction of $\widehat S$ to the eigenspace labelled by four-momentum $p$ is defined 
as $\widehat S(p)$. Since $\widehat S(p) \cdot p = 0$, it can be decomposed onto 
three orthonormal spacelike four-vectors $n_1(p), n_2(p), n_3(p)$ orthogonal to $p$, 
forming a basis of the Minkowski space with the unit vector $\hat p = p/\sqrt{p^2}$:
\begin{equation}\label{sdecomp}
  \widehat S(p) = \sum_{i=1}^3 \widehat S_i(p) n_i(p)
\end{equation}
It can be shown that the operators $\widehat S_i(p)$ form a SU(2) algebra and 
are the generators of the so-called {\em little group} of massive particles. The 
third component $\widehat S_3(p)$ can be diagonalized along with $\widehat S^2$ which 
corresponding eigenvalues $s$ and $S(S+1)$, $S$ being {\em the} spin of the 
particle so that:
\begin{equation} \label{quantiz}
 \widehat{P}\ket{p,s} = p \ket{p,s} \qquad {\rm and} \qquad 
 \widehat{S}_3(p)\ket{p,s} = s \ket{p,s} 
\end{equation}
The $\hat n_i(p)$, being orthogonal to $p$, can be written as:
\begin{equation}\label{choice}
  n_i (p) = [p] {\hat e}_i  
\end{equation}
with $\hat e_i$ the $i$-th unit space vector and $[p]$ the so-called {\em standard Lorentz 
transformation} bringing the timelike vector $p_0 = (m,0,0,0)$ into the four-momentum $p$.

The choice of $[p]$ entails a specific physical meaning of the eigenvalue $s$. 
For instance, if $\theta$,$\varphi$ are the spherical coordinates of ${\bf p}$ 
and $\xi$ the rapidity of $p$, 
$$
 [p] \equiv {\sf R}_3(\varphi) {\sf R}_2 (\theta) {\sf L}_3(\xi)
$$
where ${\sf R}_k(\psi)$ are rotations around the axis $k$ with angle $\psi$ and 
${\sf L}_k(\xi)$ a Lorentz boost along the direction $k$ with hyperbolic angle $\xi$,
is a typical choice of the standard Lorentz transformation which makes $s$ the 
helicity of the particle. Finally, if the states are normalized according to
\footnote{Throughout this paper the symbol $\varepsilon$ stands for the on-shell 
energy, that is $\varepsilon =\sqrt{ \p^2 + m^2}$}: 
\begin{equation}\label{norma}
 \braket{p,r}{q,s} = 2 \varepsilon \, \delta^3({\bf p}-{\bf q})\delta_{r s}\;.
\end{equation}
we have, for the representation of a general Lorentz transformation ${\sf \Lambda}$ 
in the Hilbert space:
\begin{equation} \label{lorentztr}
\widehat{{\sf \Lambda}}\ket{p,r} = \sum_s \ket{{\sf \Lambda}p,s} 
D^S([{\sf \Lambda}p]^{-1}{\sf \Lambda} [p])_{sr} 
\end{equation}
where $D^S$ stands for the $(2S+1)$-dimensional irreducible representation of the
proper orthocronous Lorentz group SO(1,3) (the so-called $(0,S)$ representation) 
or - in case - of its universal covering group SL(2,C). The transformation 
\be\label{wigrot}
  W({\sf \Lambda},p) = [{\sf \Lambda}p]^{-1}{\sf \Lambda} [p]
\ee
is the so-called {\em Wigner rotation}, as it leaves the unit time vector $\hat t$ 
invariant. 

The mean value of $S^\mu$ of the operator \eqref{luba} is the properly called 
{\em spin vector} (the polarization vector being the mean spin vector $S^\mu$ 
divided by $S$ so that its maximal magnitude is always 1): 
$$
  S^\mu = \Tr ( \widehat S^\mu \wrho )
$$
where $\wrho$ is the density operator of the single quantum relativistic particle
in the Hilbert space. Its restriction to the subspace of four-momentum $p$
is the {\em spin density operator} $\widehat \Theta(p)$, which is used to express the
mean value of the spin vector for a particle with momentum $p$:
\be\label{means}
  S^\mu(p) = \Tr( \widehat S^\mu \widehat \Theta(p))
\ee
The matrix:
\be\label{sdm}
 \Theta(p)_{rs} \equiv \bra{p,r} \wrho \, \ket{p,s} = \bra{p,r} \widehat \Theta(p) 
 \ket{p,s}
\ee
is the spin density matrix, in the basis labelled by the eigenvalues of $\widehat S_3(p)$.
The matrix $\Theta$, which is hermitian, positive definite and with normalized trace,
contains the maximal information about the spin state of the particle. It is dependent
on the chosen basis, yet the mean value \eqref{means} is independent thereof, that
is of the standard Lorentz transformation associated with the eigenvalue $s$.

Plugging the \eqref{sdecomp} into the \eqref{means} we get:
\begin{align}\label{meansp2}
  S^\mu(p) &= \sum_r \sum_i \bra{p,r} \widehat S_i(p) \widehat \Theta(p)
\ket{p,r} n_i(p)
  = \sum_{r,s} \sum_i \bra{p,r} \widehat S_i(p) \ket{p,s}\bra{p,s} 
 \widehat \Theta(p) \ket{p,r} n_i(p) \nonumber \\
 & = \sum_{r,s} \sum_i D^S({\sf J}^i)_{rs} \Theta(p)_{sr} n_i(p)
  = \sum_{i=1}^3 \tr (D^S({\sf J}^i) \Theta(p)) [p](\hat e_i)^\mu
= \sum_{i=1}^3 [p]^\mu_i \tr (D^S({\sf J}^i) \Theta(p)), 
\end{align}
where we have used the fact that $\widehat S_i(p)$ are the generators of the little
group SU(2) algebra in the subspace spanned by $\ket{p}$; the $D^S({\sf J}^i)$ are
the familiar matrices of the angular momentum generators for the representation with
spin $S$. The above formula can be made covariant by introducing the definition 
of angular momenta:
$$
  D^S({\sf J}^i) = -\frac{1}{2} \epsilon^{i\lambda\nu\rho} D^S(J_{\lambda\nu}) 
  \hat t_\rho 
$$ 
where $\hat t$ is the unit time vector and the tensor $J_{\lambda\nu}$ now includes
all Lorentz transformation generators, angular momentum and boosts. Since $\hat t_\rho =
\delta^0_\rho$ we can extend the sum over $i$ from 0 to 4, because of the Levi-Civita 
tensor and write:
\be\label{meansp3}
  S^\mu(p) = -\frac{1}{2} \epsilon^{\alpha\lambda\nu\rho} \hat t_\rho [p]^\mu_\alpha 
  \tr (D^S(J_{\lambda\nu}) \Theta(p))
\ee
Now:
\begin{align*}
  [p]^\mu_\alpha \epsilon^{\alpha\lambda\nu\rho} &= [p]^\mu_\alpha ([p]^\beta_\varphi 
 [p]^{-1 \lambda}_\beta) ([p]^\gamma_\chi [p]^{-1 \nu}_\gamma) ([p]^\delta_\psi 
 [p]^{-1 \rho}_\delta) \epsilon^{\alpha\varphi\chi\psi} = 
 \left( [p]^\mu_\alpha [p]^\beta_\varphi [p]^\gamma_\chi [p]^\delta_\psi 
 \epsilon^{\alpha\varphi\chi\psi} \right)
 [p]^{-1 \lambda}_\beta [p]^{-1 \nu}_\gamma [p]^{-1 \rho}_\delta \\
 &= \det[p] \epsilon^{\mu\beta\gamma\delta} \; [p]^{-1 \lambda}_\beta 
 [p]^{-1 \nu}_\gamma [p]^{-1 \rho}_\delta = \epsilon^{\mu\beta\gamma\delta}
 [p]^{-1 \lambda}_\beta [p]^{-1 \nu}_\gamma [p]^{-1 \rho}_\delta
\end{align*}
and, substituting in \eqref{meansp3}:
$$
  S^\mu(p) = -\frac{1}{2} \epsilon^{\mu\beta\gamma\delta} [p]^{-1 \rho}_\delta 
 \hat t_\rho [p]^{-1 \lambda}_\beta [p]^{-1 \nu}_\gamma \tr (D^S(J_{\lambda\nu}) \Theta(p))
$$
By definition of standard Lorentz transformation, and taking into account its 
orthogonality, we have:
$$
 [p]^{-1 \rho}_\delta \hat t_\rho = \frac{p_\delta}{m}
$$
so that the mean spin vector becomes:
$$
  S^\mu(p) = -\frac{1}{2m} \epsilon^{\mu\beta\gamma\delta} p_\delta 
  [p]^{-1 \lambda}_\beta [p]^{-1 \nu}_\gamma \tr (D^S(J_{\lambda\nu}) \Theta(p))
$$
From group representation theory we know that:
\be\label{repcon}
  [p]^{-1 \lambda}_\beta [p]^{-1 \nu}_\gamma D^S(J_{\lambda\nu}) = 
  D^S([p])^{-1} D^S( J_{\beta\gamma}) D^S([p])
\ee
hence the \eqref{meansp3} can be finally cast as:
\be\label{meansp4}
  S^\mu(p) = -\frac{1}{2m} \epsilon^{\mu\beta\gamma\delta} p_\delta 
  \tr \left( D^S([p])^{-1} D^S( J_{\beta\gamma}) D^S([p]) \Theta(p) \right)
\ee

The ``kinematic" part of the spin vector derivation for a single quantum relativistic
particle is completed. 

In a quantum field theory framework, the single-particle spin density matrix can 
be still defined with a formula which is a generalization of the \eqref{sdm}:  
\be\label{spindens}
\Theta(p)_{r s} = \frac{\Tr (\wrho \, \wadf{s}{p} \waf{r}{p} )}
{\sum_{t} \Tr (\wrho \, \wadf{t}{p} \waf{t}{p} )},
\ee
where $\wrho$ is the density operator for the Hilbert space of the field states.
The $\waf{r}{p}$ are destruction operators of the particle with momentum 
$p$ and spin state $r$. The introduction of creation and destruction operators
makes it clear that one can define a polarization of particles only when particle
is a sensible concept, that is for a non-interacting or a weakly interacting field
theory. For instance, defining a spin density matrix of quarks and gluons makes sense 
only in the perturbative limit of QCD. 

The calculation of $\Theta(p)$ is the crucial and hardest part of the procedure.
Before tackling the full quantum field theory case, one can obtain a good approximation
by using the single quantum relativistic particle formalism, which is the subject
of the next section.

\section{The single particle limit and global equilibrium factorization}
\label{single}

In the fixed-number particle formalism, the Hilbert space of quantum states is the
tensor product of single-particle Hilbert spaces. Neglecting symmetrization or 
anti-symmetrization of the states means disregarding quantum statistics effects 
and taking the limit of distinguishable particles. Moreover, if the particles are 
non-interacting, the full density operator can be written as the tensor product of 
single-particle density operators:
$$
\wrho = \otimes_i \wrho_i  
$$
We can now set out to get the spin density matrix for the general {\em global} 
equilibrium density operator $\wrho$ \cite{Becattini:2012tc,Becattini:2014yxa}:
\be\label{global}
\wrho = \frac{1}{Z} \exp \left[ -b \cdot \wP + \frac{1}{2} \varpi : \wJ \right],
\ee
where $b$ is a constant time-like four-vector and $\varpi$ a constant anti-symmetric 
tensor. In global equilibrium, the vector field:
\be\label{fourt}
  \beta^\mu = b^\mu + \varpi^{\mu\nu} x_\nu
\ee
is the four-temperature vector \cite{Becattini:2012tc} fulfilling Killing equation
and:
\be\label{thvort}
  \varpi_{\mu\nu} = -\frac{1}{2} \left( \partial_\mu \beta_\nu - \partial_\nu \beta_\mu
  \right)
\ee
is called thermal vorticity; the relation \eqref{thvort} can be taken as a definition
of thermal vorticity in non-equilibrium situation.
The operators $\wP$ and $\wJ$ in \eqref{global} are the conserved total four-momentum 
and total angular momentum-boosts, respectively. For a set of non-interacting 
distinguishable particles, we can write:
$$
\wP = \sum_{i} \wP_i, \qquad \qquad  \wJ = \sum_i \wJ_i,
$$
and consequently,
$$
 \wrho_i = \frac{1}{Z_i} \exp \left[ -b \cdot \wP_i + \frac{1}{2} \varpi : \wJ_i \right].
$$
so that the single-particle spin density matrix reads:
\be\label{spindens2}
\Theta(p)_{i \,rs} = \frac{\bra{p,r} \wrho_i \ket{p,s}}
{\sum_r \bra{p,t} \wrho_i \ket{p,t}}.
\ee

In order to calculate the right hand side of \eqref{spindens2}, one can take advantage
of a noteworthy factorization:
\be\label{factor1}
 \frac{1}{Z_i} \exp \left[ -b \cdot \wP_i + \frac{1}{2} \varpi : \wJ_i \right] 
 = \frac{1}{Z_i} \exp \left[ -\tilde b \cdot \wP_i \right] 
 \exp \left[\frac{1}{2} \varpi : \wJ_i\right],
\ee
where 
\be\label{btilde}
\tilde b_\mu = \sum_{k=0}^\infty \frac{\ii^k}{(k+1)!} \underbrace{\left( 
 \varpi_{\mu\nu_1} \varpi^{\nu_1\nu_2} \ldots \varpi_{\nu_{k-1}\nu_k} \right)}_\text{k times} 
b^{\nu_k}. 
\ee
The \eqref{factor1} is a very useful formula, whose derivation is worth being shown 
in some detail.

Let us start with the following very simple observation concerning the composition 
of translations and Lorentz transformation in Minkowski space-time. Let $x$ be a 
four-vector and apply the combination 
$$
{\sf T}(a) \, {\sf \Lambda} \, {\sf T}(a)^{-1}
$$
${\sf T}(a)$ being a translation of some four-vector $a$ and ${\sf \Lambda}$ a 
Lorentz transformation. The effect of the above combination on $x$ reads:
$$
 x \mapsto x - a \mapsto {\sf \Lambda}(x-a) \mapsto {\sf \Lambda}(x-a)+a
 = {\sf \Lambda}(x) + ({\sf I-\Lambda})(a)= {\sf T}(({\sf I-\Lambda})(a))
  ({\sf \Lambda}(x))
$$ 
Since $x$ was arbitrary, we have:
$$
 {\sf T}(a) \, {\sf \Lambda} \, {\sf T}(a)^{-1} = {\sf T}(({\sf I-\Lambda})(a)))
 {\sf \Lambda}
$$
This relation has a representation of unitary operators in Hilbert space, which can be 
written in terms of the generators of the Poincar\'e group:
\be\label{poinc1}
\exp[i a \cdot \widehat P] \exp[-\ii \phi : \wJ/2] \exp[-i a \cdot \widehat P]
= \exp[\ii(({\sf I-\Lambda})(a))\cdot \widehat{P}] \exp[-\ii \phi : \wJ/2] 
\ee
where $\phi$ are the parameters of the Lorentz transformation \footnote{Henceforth,
by $:$ we will denote a double contraction of rank 2 tensors, e.g. $\phi : \wJ 
= \phi_{\mu\nu} \wJ^{\mu\nu}$}. By taking $\phi$ infinitesimal, we can obtain a 
known relation about the effect of translations on angular momentum operators:
$$
\exp[i a \cdot \widehat P] \wJ_{\mu\nu} \exp[-i a \cdot \widehat P]
=  \widehat{\sf T}(a) \wJ_{\mu\nu} \widehat{\sf T}(a)^{-1} = \wJ_{\mu\nu}
 - a_\mu \widehat P_\nu + a_\nu \widehat P_\mu 
$$
The left hand side of (\ref{poinc1}) can now be worked out by using the above
relation:
\begin{align}\label{poinc2}
 & \exp[i a \cdot \widehat P] \exp[-\ii \phi : \wJ/2] \exp[-i a \cdot \widehat P]
 = \widehat{\sf T}(a) \exp[-\ii \phi : \wJ/2] \widehat{\sf T}(a)^{-1} \nonumber \\
& = \exp[-\ii \phi : \widehat{\sf T}(a) \wJ \widehat{\sf T}(a)^{-1}/2]
 = \exp[-\ii \phi : (\wJ - a \wedge \widehat{P})/2] = \exp[\ii \phi_{\mu\nu} a^\mu 
 \widehat{P}^\nu -\ii \phi_{\mu\nu} \wJ^{\mu\nu}/2]
\end{align}
Hence, combining (\ref{poinc2}) with (\ref{poinc1}), we have obtained the factorization:
\be\label{intermed}
\exp[\ii \phi_{\mu\nu} a^\mu \widehat{P}^\nu -\ii \phi_{\mu\nu} \wJ^{\mu\nu}/2] 
= \exp[\ii(({\sf I-\Lambda})(a))\cdot \widehat{P}] \exp[-\ii \phi : \wJ/2]
\ee
Now:
\be\label{poinc3}
 \ii({\sf I-\Lambda})(a) = \ii a - \ii 
 \sum_{k=0}^\infty \frac{(-\ii)^k}{2^k k!}(\phi: {\sf J})^k (a)
  = - \ii \sum_{k=1}^\infty \frac{(-\ii)^k}{2^k k!}(\phi: {\sf J})^k (a)
\ee
Setting:
$$
          V_\mu = \ii \phi_{\mu\nu} a^\nu
$$
and taking into account that:
$$
 ({\sf J}_{\mu\nu})^\alpha_{\, \beta} = \ii \left( \delta^\alpha_\mu g_{\nu\beta} -
 \delta^\alpha_\nu g_{\mu\beta} \right)
$$
we have:
$$
 (\phi : {\sf J})(a)_\alpha = 2 \ii \phi_{\alpha\beta}a^\beta = 2V_\alpha
$$
Therefore, the right hand side of the eq.~(\ref{poinc3}) becomes:
$$
 - \ii \sum_{k=1}^\infty \frac{(-\ii)^k}{2^k k!}(\phi: {\sf J})^k (a)
  = - \ii \sum_{k=1}^\infty \frac{(-\ii)^k}{2^{k-1} k!}(\phi: {\sf J})^{k-1}(V)
  = - \sum_{k=0}^\infty \frac{(-\ii)^{k}}{2^{k} (k+1)!}(\phi: {\sf J})^{k}(V)
$$
Finally, the eq.~(\ref{intermed}) becomes:
\be\label{intermed2}
\exp[- V \cdot \widehat{P} -\ii \phi : \wJ/2] 
= \exp[ - \tilde V (\phi) \cdot \widehat{P}] \exp[-\ii \phi : \wJ/2]
\ee
where 
$$
\tilde V (\phi) \equiv \sum_{k=0}^\infty \frac{(-\ii)^{k}}{2^{k} (k+1)!}(\phi: {\sf J})^{k}(V)
 = \sum_{k=0}^\infty \frac{1}{(k+1)!}\underbrace{\left(\phi_{\mu\nu_1} 
 \phi^{\nu_1\nu_2}\ldots \phi_{\nu_{k-1}\nu_{k}}\right)}_\text{k times} V^{\nu_k}
$$
The eq.~(\ref{intermed2}) can be read as the factorization of the exponential
of a linear combinations of generators of the Poincar\'e group. For this reason, 
it must be derivable also by using the known formulae of the factorization of the 
exponential of the sum of matrices $\exp[A+B]$ in terms of exponentials of commutators 
of $A$ and $B$. Indeed, it can be shown, by 
using the commutation relations of $\wP$ and $\wJ$, that one precisely gets the 
eq.~(\ref{intermed2}) for {\em any} vector $V$ and tensor $\phi$, either real or 
complex. Hence, the formula (\ref{intermed2}) can be applied to factorize the density 
operator \eqref{global} by setting $\phi = \ii \varpi$:
\be\label{dopfact}
\wrho = \frac{1}{Z} \exp[- b \cdot \widehat P + \varpi : \wJ/2]
 = \frac{1}{Z} \exp[-\tilde b(\varpi) \cdot \widehat P] \exp[\varpi : \wJ/2]
\ee
with:
$$
  \tilde b(\varpi) = \sum_{k=0}^\infty \frac{1}{2^{k} (k+1)!}(\varpi: {\sf J})^{k}b
 = \sum_{k=0}^\infty \frac{\ii^k}{(k+1)!}\underbrace{\left(\varpi_{\mu\nu_1} 
 \varpi^{\nu_1\nu_2}\ldots \varpi_{\nu_{k-1}\nu_{k}}\right)}_\text{k times} b^{\nu_k}
$$
We have thus proved the formula \eqref{factor1}.

The factorization of the density operator in eq.~\eqref{factor1} can now be applied
to calculate the spin density matrix in eq.~\eqref{spindens2}. The momentum-dependent
factor $\exp(-\tilde b \cdot p)$ cancels out in the ratio, and one is left with:
$$
\Theta(p)_{rs} = \frac{\bra{p,r} \exp[\varpi : \wJ/2] \ket{p,s}}
{\sum_t \bra{p,t} \exp[\varpi : \wJ/2] \ket{p,t}}.
$$
To derive its explicit form, we use an analytic continuation; namely, we first 
determine $\Theta(p)$ for imaginary $\varpi$ and then continue the function to 
real values. If $\varpi$ is imaginary, $\exp[\varpi:\wJ/2] \equiv \wLa$ is just a 
unitary representation of a Lorentz transformation, and then one can use known 
relations of Poincar\'e group representations \cite{Weinberg:1995mt} to obtain:
\be\label{spindens3}
\Theta(p)_{rs} = \frac{\bra{p,r} \wLa \ket{p,s}}
{\sum_t \bra{p,t} \wLa \ket{p,t}} = 
\frac{2\varepsilon\delta^3({\bf p} - {\bf \Lambda}(p)) W({\sf \Lambda},p)_{rs}}
{2\varepsilon\delta^3({\bf p} - {\bf \Lambda}(p)) \sum_t W({\sf \Lambda},p)_{tt}}
\ee
where ${\bf \Lambda}(p)$ stands for the spacial part of the four-vector ${\sf \Lambda}(p)$ 
and $W({\sf \Lambda},p)$ is the Wigner rotation defined in the eq.~\eqref{wigrot}. 
We thus have:
$$
  \Theta(p)_{rs} = \frac{D^S([p]^{-1} {\sf \Lambda} [p])_{rs}}
  {\tr (D^S({\sf \Lambda}))},
$$
which seems to be an appropriate form to be analytically continued to real $\varpi$. 
However, the above form is not satisfactory yet as the continuation to real $\varpi$, 
that is:
$$
  D^S({\sf \Lambda}) =  \exp\left[- \frac{\ii}{2} \varpi : \Sigma_S\right] 
  \to \exp\left[\frac{1}{2} \varpi : \Sigma_S\right]
$$
where $\Sigma_S =  D^S(J)$ is the matrix representing the generators, does not give 
rise to a hermitian matrix for $\Theta(p)$ as it should. This problem can be fixed 
by taking into account that $W(p)$ is the representation of a rotation, hence 
unitary. We can thus replace $W(p)$ with $(W(p) + W(p)^{-1\dagger})/2$ in (\ref{spindens3}) 
and, by using the property of of SL(2,C) representations $D^S(A^\dagger) = D^S(A)^\dagger$ 
\cite{Moussa:1966gjd} we obtain:
$$
  \Theta(p) = \frac{D^S([p]^{-1} {\sf \Lambda} [p])+
   D^S([p]^{\dagger} {\sf \Lambda}^{-1 \dagger}[p]^{-1\dagger})}
   {\tr (D^S({\sf \Lambda})+D^S({\sf \Lambda})^{-1 \dagger})},
$$
which will give a hermitian result because the analytic continuation of 
$\Lambda^{-1\dagger}$ reads \footnote{Note that the Lorentz transformations in 
Minkowski space-time and their 
counterparts of the fundamental $(0,1/2)$ representation of the SL(2,C) group are 
henceforth identified. Particularly, the standard Lorentz transformation $[p]$ 
will indicate either a SO(1,3) transformation or a SL(2,C) transformation.}:
$$
  D^S({\sf \Lambda}^{-1\dagger}) \to \exp\left[\frac{1}{2} \varpi : \Sigma_S^\dagger\right].
$$
Altogether, the final expression of the spin density matrix reads:
\be\label{spindensf}
  \Theta(p) = \frac{D^S([p])^{-1} \exp[(1/2) \varpi : \Sigma_S] D^S([p])+ 
   D^S([p])^{\dagger} \exp[(1/2) \varpi : \Sigma^\dagger_S] D^S([p])^{-1\dagger})}
   {\tr (\exp[(1/2) \varpi : \Sigma_S] + \exp[(1/2) \varpi : \Sigma_S^\dagger])},
\ee
which is manifestly hermitian. 

The equation (\ref{spindensf}) can be further developed. By using the \eqref{repcon},
we have:
$$
  D^S([p])^{-1} \exp \left[ \frac{1}{2} \varpi : \Sigma_S \right] D^S([p]) = 
   \exp\left[ \frac{1}{2} \varpi^{\mu\nu} D^S([p])^{-1} \Sigma_{S \, \mu\nu} D^S([p]) \right] =  
   \exp\left[ \frac{1}{2} \varpi^{\mu\nu} [p]^{-1\alpha}_\mu [p]^{-1^\beta}_\nu 
  \Sigma_{S \, \alpha\beta}
   \right].
$$
which applies to the original SO(1,3) matrices too. So, if we apply the Lorentz 
transformation $[p]$ to the tensor $\varpi$:
\be\label{varpiboost}
  \varpi^{\mu\nu} [p]^{-1\alpha}_\mu [p]^{-1 \beta}_\nu = \varpi^{\alpha\beta}_{*}(p)
\ee
we realize that $\varpi_*^{\alpha\beta}$ are the components of the thermal vorticity tensor 
in the rest-frame of the particle with four-momentum $p$. Note that these components are obtained 
by back-boosting with $[p]$ (which in fact is not a pure Lorentz boost in the helicity 
scheme as it includes a rotation). Finally, the equation (\ref{spindensf}) becomes:
\be\label{spindensf2}
  \Theta(p) = \frac{D^S(\exp[(1/2) \varpi_*(p) : \Sigma_S])+ D^S(\exp[(1/2) \varpi_*(p) : \Sigma^\dagger_S])}
   {\tr (\exp[(1/2) \varpi : \Sigma_S] + \exp[(1/2) \varpi : \Sigma_S^\dagger])}.
\ee
The thermal vorticity $\varpi$ is usually $\ll 1$; in this case, the spin density matrix can be
expanded in power series around $\varpi=0$. Taking into account that $\tr(\Sigma_S)=0$, we have:
$$
  \Theta(p)_{rs} \simeq \frac{\delta_{rs}}{2S+1} + \frac{1}{4(2S+1)} 
  \varpi_*(p)^{\alpha\beta}(\Sigma_{S{\alpha\beta}}+\Sigma_{S{\alpha\beta}}^\dagger)_{rs}
$$
to first order in $\varpi$. Now the $\Sigma_S$ matrices can be decomposed into 
representations of angular momentum and boosts:
$$
  \Sigma_{S\mu\nu} = D^S(J_{\mu\nu}) = \epsilon_{\mu\nu\rho\sigma} D^S({\sf J}^\rho) 
  \hat t^\sigma - D^S({\sf K}_\mu) \hat t_\nu + D^S({\sf K}_\nu) \hat t_\mu 
$$
and taking into account that the $D^S({\sf J}^i)$ are hermitian while $D^S({\sf K}^i)$ 
are anti-hermitian, we find:
\be\label{spindensexp}
  \Theta(p)_{rs} \simeq \frac{\delta_{rs}}{2S+1} + \frac{1}{2(2S+1)} 
  \varpi_*(p)^{\alpha\beta} \epsilon_{\alpha\beta\rho\nu} D^S({\sf J}^\rho)^r_{r'} 
  \hat t^\nu.
\ee
By plugging \eqref{spindensexp} into \eqref{meansp3}, we get:
\begin{align}\label{vortspin}
 S^\mu(p) &= [p]^\mu_\kappa \frac{1}{2(2S+1)} \varpi_*(p)^{\alpha\beta} 
  \epsilon_{\alpha\beta\rho\nu} \tr \left( D^S({\sf J}^\rho) D^S({\sf J}^\kappa) \right) 
  \hat t^\nu \nonumber \\
  & = - \frac{1}{2(2S+1)}\frac{S(S+1)(2S+1)}{3} [p]^\mu_\kappa \varpi_*(p)^{\alpha\beta} 
  \epsilon_{\alpha\beta\rho\nu} g^{\rho \kappa} \hat t^\nu \nonumber \\ 
  & = - \frac{1}{2}\frac{S(S+1)}{3} [p]^\mu_\rho \varpi_*(p)_{\alpha\beta} 
  \epsilon^{\alpha\beta\rho\nu} \hat t_\nu = - \frac{1}{2m}\frac{S(S+1)}{3} \varpi_{\alpha\beta} 
  \epsilon^{\alpha\beta\mu\nu} p_\nu,
\end{align}
where, in the last equality, we have boosted the vector to the laboratory frame by using
the Eq.~\eqref{varpiboost}. 

For a fluid made of distinguishable particles at local thermodynamic equilibrium, 
the thermal vorticity $\varpi$ is promoted to a function of space and time, so 
that the expression \eqref{vortspin} gives rise to the integral average:
\be\label{formula}
 S^\mu(p)= - \frac{1}{2m}\frac{S(S+1)}{3} \epsilon^{\mu\alpha\beta\nu} p_\nu
 \frac{\int_\Sigma d\Sigma_\lambda p^\lambda f(x,p) \varpi_{\alpha\beta}(x)}
 {\int_\Sigma d\Sigma_\lambda p^\lambda f(x,p)}
\ee
with $f(x,p)$ the distribution function and $\Sigma$ is a 3D hypersurface from 
where particles are emitted. The latter is basically the same formula 
obtained in refs.~\cite{Becattini:2013fla,Becattini:2016gvu} and should apply to 
particles with any spin.

\section{The covariant Wigner function}
\label{wigner}

We now turn to the general quantum field formula of the spin density matrix, the
equation \eqref{spindens}. To develop this expression, we need to introduce an 
important quantity: the covariant Wigner operator. We will do this first for the 
scalar field, where the spin plays no role, and later for the Dirac field.

\subsection{The scalar field}

The covariant Wigner operator is defined as a Fourier transform of the two-point
function of the quantum field:
\be\label{wscalar}
 \wW (x,k) = \frac{2}{(2\pi)^4} \int \di^4 y \; :\wpsi^\dagger(x+y/2) \wpsi(x-y/2): 
 \e^{-\ii y \cdot k}
\ee
where $:$ stands for the normal ordering of creation and destruction operators; the 
appearance of a normal ordering implies that this definition is suitable for a free 
field or a field interacting with an external field. Even if \eqref{wscalar} is not, 
strictly speaking, a local operator (it depends on the field in two points), its 
quasi-locality makes it a suitable tool to deal with local thermodynamic equilibrium 
in quantum statistical mechanics. Besides, its mean value, namely the covariant Wigner 
function:
\be\label{wscalar2}
  W(x,k) = \Tr (\wrho \; \wW (x,k))
\ee
where $\wrho$ is the density operator, is an indespensable tool to reckon quantum 
corrections to classical kinetic theory \cite{DeGroot:1980dk}. From the equation 
\eqref{wscalar2} it turns out that the covariant Wigner function is real, but it 
does not need to be positive definite. Inserting in the eq.~\eqref{wscalar} the 
free scalar field expansion in plane waves: 
\be\label{field}
 \wpsi(x) = \frac{1}{(2\pi)^{3/2}} \int \frac{\di^3 \p}{2\varepsilon} \; 
 \e^{-\ii p \cdot x} \wa{p} + \e^{\ii p \cdot x} \wbd{p}
\ee
$a_p,b_p$ being destruction operators of particles with four-momentum $p$ normalized
so as to:
\be\label{acomm}
  [\wa{p},\wad{p'}] = 2 \, \varepsilon \, \delta^3({\bf p}-{\bf p}')
\ee
we get, for the covariant Wigner function:
\begin{align}\label{wignerf}
 W(x,k) &= \frac{2}{(2\pi)^{7}} \int \frac{\di^3 \p}{2\varepsilon} \frac{\di^3 \p^\prime}
 {2\varepsilon^\prime}\int \di^4 y \; \e^{-\ii y \cdot (k - (p+p^\prime)/2)} \e^{\ii (p-p^\prime)\cdot x}
 \media{\wad{p}\wa{p^\prime}} + \e^{-\ii y \cdot (k + (p+p^\prime)/2)} \e^{-\ii (p-p^\prime)\cdot x}
 \media{\wbd{p^\prime}\wb{p}} \nonumber \\
 &+ \e^{-\ii y \cdot (k - (p-p^\prime)/2)} \e^{\ii (p + p^\prime)\cdot x}
 \media{\wad{p}\wbd{p^\prime}} + \e^{-\ii y \cdot (k + (p - p^\prime)/2)} \e^{-\ii (p+p^\prime)\cdot x}
 \media{\wb{p}\wa{p^\prime}} \nonumber \\
 &= \frac{2}{(2\pi)^{7}} \int \frac{\di^3 \p}{2\varepsilon} \frac{\di^3 \p^\prime}
 {2\varepsilon^\prime}\int \di^4 y \; \e^{-\ii y \cdot (k - (p+p^\prime)/2)} \e^{\ii (p-p^\prime)\cdot x}
 \media{\wad{p}\wa{p^\prime}} + \e^{-\ii y \cdot (k + (p+p^\prime)/2)} \e^{\ii (p-p^\prime)\cdot x}
 \media{\wbd{p}\wb{p^\prime}} \nonumber \\
 &+ \e^{-\ii y \cdot (k - (p-p^\prime)/2)} \e^{\ii (p + p^\prime)\cdot x}
 \media{\wad{p}\wbd{p^\prime}} + \e^{-\ii y \cdot (k - (p - p^\prime)/2)} \e^{-\ii (p+p^\prime)\cdot x}
 \media{\wb{p^\prime}\wa{p}} \nonumber \\
 &= \frac{2}{(2\pi)^{3}} \int \frac{\di^3 \p}{2\varepsilon} \frac{\di^3 \p^\prime}
{2\varepsilon^\prime} \e^{\ii (p-p^\prime)\cdot x} \left[ \delta^4(k - (p+p^\prime)/2) 
 \media{\wad{p}\wa{p^\prime}} + \delta^4(k + (p+p^\prime)/2) \media{\wbd{p}\wb{p^\prime}} \right] \nonumber \\
 &+ \delta^4 (k - (p - p^\prime)/2) \left[ \e^{\ii (p + p^\prime)\cdot x} 
\media{\wad{p}\wbd{p^\prime}} +  \e^{- \ii (p + p^\prime)\cdot x} \media{\wb{p^\prime}\wa{p}} \right]
\end{align}
where $\langle \; \rangle$ stands for the mean $\Tr (\wrho \;)$. In the above equalities,
we have taken advantage of the symmetric integration in the variables $p,p^\prime$.

The expression (\ref{wignerf}) makes it apparent that the variable $k$ of the Wigner 
function $W(x,k)$ is {\em not} on-shell, i.e. $k^2 \ne m^2$ even in the free case. 
This makes the definition of a particle distribution function $f(x,p)$ \`a la Boltzmann 
not straighforward in a quantum relativistic framework. In the book by De Groot 
\cite{DeGroot:1980dk} it is shown that a distribution function can be defined in 
the limit of slowly varying $W(x,k)$ on the microscopic scale of the Compton wavelength. 
In fact, we will show that in the case of the free scalar field a distribution 
function can be defined without introducing such an approximation.
From equation~(\ref{wignerf}), we can infer that $W$ is made up of three terms which
can be distinguished for the characteristic of $k$. For future time-like $k = (p+p^\prime)/2$, 
only the first term involving particles is retained; for past time-like $k= - (p+p^\prime)/2$, 
only the second term involving antiparticles; finally, for space-like $k = (p-p^\prime)/2$ 
the last term with the mean values of two creation/destruction operators is retained. 
In symbols: 
\be\label{wdecomp}
    W(x,k ) = W(x,k)\theta(k^2)\theta(k^0) + W(x,k) \theta(k^2) \theta(-k^0) +
    W(x,k) \theta(-k^2) \equiv  W_+(x,k) + W_-(x,k) + W_S(x,k)
\ee

Local operators quadratic in the field can be expressed as four-dimensional 
integrals over $k$ of the covariant Wigner function (\ref{wscalar}). For instance,
the mean value of the conserved current of the scalar field is defined as 
\cite{DeGroot:1980dk}:
\be\label{wcurr}
  j^\mu(x) =  \ii \media{: \wpsi^\dagger(x) \codevmu \wpsi(x):} = 
  \int \di^4 k \; k^\mu W(x,k)
\ee
By using the decomposition \eqref{wdecomp} for $W(x,k)$, the current can be written 
as the sum of three terms. The particle term, by using the \eqref{wignerf}, reads:
\begin{align}\label{current}
 j^\mu_+(x) &= \frac{1}{(2\pi)^{3}} \int \frac{\di^3 \p}{\varepsilon} 
  \frac{\di^3 \p^\prime}{2\varepsilon^\prime} \;  \frac{(p + p^\prime)^\mu}{2} \, 
   \e^{\ii (p-p^\prime)\cdot x} \media{\wad{p}\wa{p^\prime}} \nonumber \\
 &= \int \frac{\di^3 \p}{\varepsilon} p^\mu {\rm Re} \left(\frac{1}{(2\pi)^3} 
  \int \frac{\di^3 \p^\prime}{2\varepsilon^\prime} \; \e^{\ii (p-p^\prime)\cdot x} 
  \media{\wad{p}\wa{p^\prime}} \right)
\end{align}
To obtain the last expression, we have taken advantage of the hermiticity of the density
operator, implying:
$$
  \media{\wad{p}\wa{p^\prime}} = \media{\wad{p^\prime}\wa{p}}^*
$$
which makes it possible to swap the integration variables $p$ and $p^\prime$. 
The formula (\ref{current}) brings out a function that we can properly identify
as the particle distribution function or phase space density, as the real part 
of a {\em complex distribution function} $f_c(x,p)$:
\be\label{phspace}
 f(x,p) = {\rm Re} f_c(x,p)
\ee
where
\be\label{phspacec}
 f_c(x,p) = \frac{1}{(2\pi)^3} \int \frac{\di^3 \p^\prime}{2\varepsilon^\prime} 
   \; \e^{\ii (p-p^\prime)\cdot x} \media{\wad{p}\wa{p^\prime}}
\ee
Similarly, the antiparticle term in the current leads to a distribution function 
$\bar f(x,p)$ which is obtained from (\ref{phspacec}) replacing 
$\media{\wad{p}\wa{p^\prime}}$ with $\media{\wbd{p}\wb{p^\prime}}$. Finally, the 
last term in the current, can be written, by using (\ref{wcurr}) and (\ref{wignerf}):
\begin{align*}
  j^\mu_S(x) &= \frac{1}{(2\pi)^{3}} \int \frac{\di^3 \p}{\varepsilon} 
  \frac{\di^3 \p^\prime}{2\varepsilon^\prime} \;  (p - p^\prime)^\mu \, 
   {\rm Re} \left( \e^{\ii (p+p^\prime)\cdot x} \media{\wad{p}\wbd{p^\prime}} \right)
   \\
   &= \frac{1}{(2\pi)^{3}} \int \frac{\di^3 \p}{\varepsilon} 
  \frac{\di^3 \p^\prime}{2\varepsilon^\prime} \;  p^\mu \, 
   {\rm Re} \left[ \e^{\ii (p+p^\prime)\cdot x} \left( 
   \media{\wad{p}\wbd{p^\prime}} - \media{\wad{p^\prime}\wbd{p}} \right) \right]
\end{align*}
where, again, we have used the hermiticity of the density operator. Thereby, we 
could define a mixed distribution function $g_c$:
$$
 g_c(x,p) = \frac{1}{(2\pi)^3} \int \frac{\di^3 \p^\prime}{2\varepsilon^\prime} 
   \; \left[ \e^{\ii (p+p^\prime)\cdot x} \left( 
   \media{\wad{p}\wbd{p^\prime}} - \media{\wad{p^\prime}\wbd{p}} \right) \right]
$$
and write the whole current as: 
$$
 j^\mu(x) =  \int \di^4 k \; k^\mu W(x,k) = {\rm Re}
 \int \frac{\di^3 \p}{2\varepsilon} p^\mu \left[ f_c(x,p) - {\bar f}_c(x,p) + g_c(x,p) 
  \right]  
= \int \frac{\di^3 \p}{2\varepsilon} p^\mu \left[ f(x,p) - {\bar f}(x,p) + g (x,p)
  \right] 
$$
If $g_c=0$, which is the most common case, the current can be formally written as 
the familiar relativistic kinetic formula, that is an integral over on-shell 
four-momenta of the four-momentum vector multiplied by on-shell phase space densities.

In general, an algebraic relation between e.g. $W_+(x,k)$ and $f(x,p)$ does not exist. 
Nevertheless, interesting integral relations between them can be obtained. It is
not hard to show that, integrating the \eqref{wignerf} in $k$, one gets:
\be\label{intrel1}
  \int \di^4 k \; W(x,k) = \int \frac{\di^3 \p}{\varepsilon} \left(
   f_c(x,p) + \bar f_c(x,p) + g_c (x,p) \right) 
   = \int \frac{\di^3 \p}{\varepsilon} \left( f(x,p) + \bar f(x,p) + g(x,p) \right)
\ee
where the last equality follows from the vanishing of the imaginary part of the 
integral. Furthermore, if we integrate the time component of the particle current 
\eqref{current} over the hypersurface $t=const$ one retrieves the total number of 
particles:
$$
 N= \int \di^3 \x \;  j^0_+(x) = \int \di^3 \x \frac{\di^3 \p}{\varepsilon} 
 \varepsilon f(x,p) = \int \di^3 \p \int \di^3 \x \; f(x,p)
$$
the last expression confirms that $f(x,p)$ is the actual density of particles 
in phase space. Furthermore, according to the eq.~\eqref{phspacec}:
$$ 
   \frac{\di N}{\di^3 p} = \int \di^3 \x \; f(x,p) = \frac{1}{2\varepsilon}
  \media{\wad{p}\wa{p}}
$$
which is the expected relation between the particle density in momentum space 
in view of the \eqref{acomm}.

\subsection{The Dirac field}
\label{dirac}

A similar connection can be built up for the spin 1/2 particles and the Dirac field.
In this case, the covariant Wigner operator is a $4 \times 4$ spinorial matrix 
\footnote{It should be reminded that the normal ordering for fermion fields involves 
a minus sign for each permutation, e.g. $:a a^\dagger: = - a^\dagger a$. Therefore, 
taking into account anticommutation relations, for fields $:\Psi_A(x)\Psibar_B(y): 
= -:\Psibar_B(y) \Psi_A(x):$}:
\bea\label{wigdirop}
   \wW(x,k)_{AB} &=& - \frac{1}{(2\pi)^4} \int \di^4 y \; \e^{-\ii k \cdot y}
    : \Psi_A (x-y/2) \Psibar_B (x+y/2) : \nonumber \\
   &=& \frac{1}{(2\pi)^4} \int \di^4 y \; \e^{-\ii k \cdot y} : 
   \Psibar_B (x+y/2) \Psi_A (x-y/2):  
\eea
$\Psi$ being the Dirac field:
\be\label{dirf}
  \Psi_A(x) = \sum_r \frac{1}{(2\pi)^{3/2}} \int \frac{\di^3 \p}{2 \varepsilon} 
  \; \waf{r}{p} u_r(p)_A e^{-\ii p \cdot x} + \wbdf{r}{p} 
  v_r(p)_A e^{\ii p \cdot x}
\ee
In the equation \eqref{dirf}, the creation and destruction operators are normalized 
according to the \eqref{acomm} with the anticommutator replacing the commutator,
and $u_r(p),v_r(p)$ are the spinors of free particles and antiparticles 
in their polarization state $r$ (usually a helicity or third spin component) 
normalized so as to $\bar u_r u_s = 2m \delta_{rs}$, $\bar v_r
v_s = -2m \delta_{rs}$. The covariant Wigner function is, again, the mean value
of the Wigner operator in \eqref{wigdirop}. The definition \eqref{wigdirop} has 
to be modified in full spinor electrodynamics \cite{Vasak:1987um} to preserve gauge 
invariance, but this can be neglected for the scope of this work. Because of the 
Dirac equation, the Wigner operator solves the equation:
$$
  \left( m - \slashed{k} - \frac{i}{2} \slashed{\partial} \right) \wW(x,k) = 0
$$

By plugging the \eqref{dirf} into the \eqref{wigdirop}, we obtain:
\begin{align}\label{wignerdop}
 \wW(x,k)_{AB} = & \sum_{r,s} 
 \frac{1}{(2\pi)^3} \int \frac{\di^3 \p}{2\varepsilon} \frac{\di^3 \p^\prime}
 {2\varepsilon^\prime} \e^{-\ii (p-p^\prime)\cdot x} \left[ \delta^4(k - (p+p^\prime)/2) 
  \wadf{s}{p^\prime} \waf{r}{p} u_r(p)_A \bar u_s(p')_B \right. \nonumber \\
 - & \left. \delta^4(k + (p+p^\prime)/2) \wbdf{r}{p} \wbf{s}{p^\prime} 
  v_r(p)_A \bar v_s(p')_B \right] \nonumber \\
&- \delta^4 (k - (p - p^\prime)/2) \left[ \e^{-\ii (p + p^\prime)\cdot x} 
  \waf{r}{p} \wbf{s}{p^\prime} u_r(p)_A \bar v_s (p^\prime)_B  +  
  \e^{\ii (p + p^\prime)\cdot x} v_r(p^\prime)_A \bar u_s (p)_B  
   \wbdf{r}{p^\prime} \wadf{s}{p} \right]
\end{align}
It can be seen that in the fermion case, the covariant Wigner operator can be
split into future time-like (particle), past time-like (antiparticle) and space-like 
parts corresponding to the three terms of the right hand side, just as for the
scalar field:
$$
   \wW(x,k) = \wW(x,k) \theta(k^2)\theta(k^0) +\wW(x,k)_{AB} 
   \theta(k^2) \theta(-k^0) +
    \wW(x,k) \theta(-k^2) \equiv \wW_+(x,k) +\wW_-(x,k) +
   \wW_S(x,k)
$$
Yet, unlike for the scalar field, we cannot identify a distribution function in 
phase space by integrating the covariant Wigner function in $k$. We can make this 
clear by calculating the mean current of the Dirac field (without vacuum contribution)
which is obtained from the Wigner function through the formula \cite{DeGroot:1980dk}:
$$
 j^\mu(x) = \langle : \Psibar (x) \gamma^\mu \Psi(x) : \rangle = 
  \int \di^4 k \; \tr (\gamma^\mu W(x,k))
$$
We confine ourselves to the particle term of the \eqref{wignerdop}, which, once 
fed into the above formula yields:
\be\label{partcurr}
 j_+^\mu(x) = \int \di^4 k \; \tr (\gamma^\mu W(x,k)_+) 
  = \sum_{r,s} \frac{1}{(2\pi)^3} \int \frac{\di^3 \p}{2\varepsilon} 
 \frac{\di^3 \p^\prime} {2\varepsilon^\prime} \e^{-\ii (p-p^\prime)\cdot x} 
 \media{\wadf{s}{p^\prime}\waf{r}{p}} \bar u_s (p') \gamma^\mu u_r(p)
\ee
Unlike in the \eqref{current} we cannot factorize the momentum integration because
the spinors $u$ have different momenta as argument. It thus follows that a reasonable 
definition of a particle distribution function with spin indices $f(x,p)_{rs}$
is precluded, except in the limit of very slow variation of the Wigner function
as derived in ref.~\cite{DeGroot:1980dk}. 

Notwithstanding, it is possible to establish an exact relation between the density 
of particles in momentum space and the covariant Wigner function. First of all, 
it can be shown, from \eqref{partcurr}, that:
$$
  \partial_\mu j_+^\mu = 0
$$
and likewise for $j_-^\mu$, taking into account that the spinors $u(p)$ and $\bar u(p)$ 
fulfill the equations:
$$
   (\slashed p - m ) u(p) = 0 \qquad \qquad  \bar u(p) (\slashed p -m) = 0
$$
If the divergence of $j_+$ vanishes, we can integrate the particle current \eqref{partcurr} 
over an arbitrary 3D space-like hypersurface to get a constant particle number,
provided that boundary fluxes vanish. For instance, we can integrate $j^0_+$ over 
the hypersurface $t=const$ to get:
\begin{align*}
 N &= \int \di^3 \x \; j_+^0(x) = \int \di^3 \x \int \di^4 k \; \tr (\gamma^0 W(x,k)_+) 
  = \sum_{r,s} \int \frac{\di^3 \p}{2\varepsilon} 
  \frac{\di^3 \p^\prime} {2\varepsilon^\prime} \delta^3 ({\bf p} - {\bf p}^\prime) 
 \media{\wadf{r}{p}\waf{s}{p^\prime}} \bar u_s (p') \gamma^0 u_r(p)
 \\
 &= \sum_{r,s} \int \frac{\di^3 \p}{4\varepsilon^2}  
 \media{\wadf{r}{p}\waf{s}{p}} \bar u_s (p) \gamma^0 u_r(p)
 = \sum_{r,s} \int \frac{\di^3 \p}{4\varepsilon^2}  
 \media{\wadf{r}{p}\waf{s}{p}} 2 \varepsilon \delta_{r,s}
 = \int \frac{\di^3 \p}{2\varepsilon} \sum_{r} \media{\wadf{r}{p}\waf{r}{p}} 
\end{align*}
whence we obtain the particle density in momentum space, as expected:
\be\label{momdens}
 \frac{\di N}{\di^3 p} = \frac{1}{2 \varepsilon} \sum_r 
 \media{\wadf{r}{p}\waf{r}{p}} 
\ee

An important feature of the Wigner operator of free fields is that integrating it
over a 3D hypersurface, $k$ becomes an on-shell vector. Indeed, from \eqref{wignerdop}:
$$
 k^\mu \partial_\mu \wW_\pm(x,k) = k^\mu \partial_\mu \wW_S(x,k) = 0
$$
because, in taking the derivative $k\cdot \partial$ the factor $(p-p^\prime)\cdot 
(p+p^\prime) =0$ is generated in all of the terms; the same applies to the Wigner
operator of the scalar field. Therefore, provided that suitable boundary conditions 
are fulfilled, the integral over a space-like 3D hypersurface: 
$$
  \int_\Sigma \di \Sigma_\mu k^\mu \wW (x,k) 
$$
is independent of the hypersurface $\Sigma$. Thus, we can choose $\Sigma$ at the 
hyperplane $t=0$ and obtain, from \eqref{wignerdop}:
\begin{align}\label{wignerint}
 & \int_{t=0} \di \Sigma_\mu k^\mu \wW(x,k) = k^0 \int \di^3 \x \; 
 \wW(x,k) =
 \sum_{r,s} k^0 \int \frac{\di^3 \p}{2\varepsilon} \frac{\di^3 \p^\prime}
 {2\varepsilon^\prime} \delta^3({\bf p} - {\bf p}^\prime) \left[ \delta^4(k - p) 
 \wadf{s}{p} \waf{r}{p} u_r(p) \bar u_s(p) \right. 
 \nonumber \\
 - & \left. \delta^4(k + p) \wbdf{r}{p} \wbf{s}{p} 
  v_r(p)_A \bar v_s(p)_B \right] + 
 \delta(k^0) \delta^3 ({\bf k} - {\bf p}) \left[ \delta^3({\bf p}+{\bf p}^\prime)  
  \waf{r}{p} \wbf{s}{p^\prime} u_r(p) \bar v_s (p^\prime)  +  
  v_r(p^\prime) \bar u_s (p)  
  \wbdf{r}{p^\prime} \wadf{s}{p} \right] \nonumber \\
 &= \sum_{r,s} k^0 \int \frac{\di^3 \p}{4\varepsilon^2} 
 \left[ \delta^4(k - p) \wadf{s}{p} \waf{r}{p} 
 u_r(p) \bar u_s(p) - \delta^4(k + p) \wbdf{r}{p} \wbf{s}{p} 
  v_r(p) \bar v_s(p) \right] \nonumber \\
&= \sum_{r,s} \frac{1}{2} \delta(k^2-m^2) \left[ \theta(k^0)
 \wadf{s}{k} \waf{r}{k} u_r(k) \bar u_s(k) + \theta(-k^0) 
 \wbdf{r}{-k} \wbf{s}{-k} v_r(-k)_A \bar v_s(-k)_B \right]  \; ;
\end{align}
note that the mixed term vanished because of the factor $k^0 \delta(k^0)$. 
The last obtained result proves the above statement, that is the argument $k$ of 
the integral over a 3D spacelike hypersurface $\Sigma$ is an on-shell four vector
and it nicely separates particle-antiparticle contribution. Indeed, the on-shell
operators $\widehat w_\pm$ can be defined such that:
\be\label{reducedwig}
  \frac{1}{2 \varepsilon_k} \delta(k^0 - \varepsilon_k) \widehat w_+(k) = 
  \int \di \Sigma_\mu k^\mu \wW_+(x,k) \qquad \qquad 
  \frac{1}{2 \varepsilon_k} \delta(k^0 + \varepsilon_k) \widehat w_-(k) = 
  \int \di \Sigma_\mu k^\mu \wW_-(x,k)
\ee
so that, by comparing with the \eqref{wignerint}:
\be\label{wignerint2}
 \ww_+(k) = \sum_{r,s} \frac{1}{2} 
 \wadf{s}{k} \waf{r}{k} u_r(k) \bar u_s(k)
\ee
with $k$ on-shell. A similar equation can be established for $\wW_-$ and 
antiparticles. Note that, from \eqref{wignerint2}:
\be\label{diracw}
    (\slashed k - m) \ww_+(k) = \ww_+(k) (\slashed k - m) = 0
\ee
Now, multiplying the \eqref{wignerint2} by $\bar u_r(k)$ to the left and $u_s(k)$
to the right, and keeping in mind the normalization of the spinors $u$, we get:
\be\label{wignerint3}
 \bar u_r(k) \ww_+(k) u_s(k) = 2 m^2 \wadf{s}{k} \waf{r}{k} 
\ee
This formula will be used in the next section to express the spin density matrix. 
Now, setting $r=s$, summing over $r$ and taking the mean value with the suitable
density operator, we obtain:
$$
 \sum_r \bar u_r(k) w_+(k) u_r(k) = \tr \left( w_+(k) \sum_r u_r(k) \bar u_r (k) \right) 
 = 2 m^2 \sum_r \media{\wadf{r}{k} \waf{r}{k}} = 4 m^2 \varepsilon \frac{\di N}{\di^3 {\rm k}} 
$$
where we have used the \eqref{momdens}. Since:
$$
  \sum_r u_r(k) \bar u_r(k) = \slashed{k} + m
$$
we have:
$$
 \varepsilon \frac{\di N}{\di^3 {\rm k}} = \frac{1}{4m^2} \tr \left( w_+(k) (\slashed k + m)
\right)         
$$
and, by using the \eqref{diracw}, we finally obtain the sought relation between the 
momentum spectrum and the covariant Wigner function:
\be\label{momdens2}
 \varepsilon \frac{\di N}{\di^3 {\rm k}} = \frac{1}{2m} \tr \, w_+(k) = 
  \frac{\varepsilon}{m} \int \di k^0 \int \di \Sigma_\mu \, k^\mu \; \tr \, W_+(x,k)    
\ee
%

\section{Fermion polarization and the covariant Wigner function}
\label{polwig}

We are now in a position to derive an exact formula connecting the Wigner function
to the spin density matrix and the spin vector for spin 1/2 fermions. 
The derivation of $\Theta(p)$ for particles is now a straightforward consequence 
of its definition \eqref{spindens} and of the \eqref{wignerint3}:
$$
 \Theta(p)_{rs} = \frac{\bar u_r(p) w_+(p) u_s(p)}
 {\sum_t \bar u_t(p) w_+(p) u_t(p)}
$$
This formula can be also written in an expanded form by using the actual covariant
Wigner function by using the \eqref{reducedwig} and taking advantage of the cancellation
of the Dirac deltas in the ratio:
\be\label{spindensw}
 \Theta(p)_{r s} = 
 \frac{\int \di \Sigma_\mu p^\mu \bar u_r(p) W_+(x,p) u_s(p)}
 {\sum_t \int \di \Sigma_\mu p^\mu \bar u_t(p) W_+(x,p) u_t(p)}
\ee
keeping in mind that $p$ is on-shell because of the integration. As we have emphasized 
in the previous section, as long as one deals with free fields, the integration 
hypersurface is arbitrary, and this will be important for the use of \eqref{spindensw}
in relativistic heavy ion collisions. The above matrix can be written in a more 
compact way by introducing $4 \times 2$ spinorial matrices $U$ (and corresponding 
$2 \times 4$ $\bar U$) such that $U_{A,r}(p) = u_r(p)_A$:
\be\label{spindensw2}
 \Theta(p) = \frac{\int \di \Sigma_\mu p^\mu \bar U(p) W_+(x,p) U (p)}
 {\tr_2 \int \di \Sigma_\mu p^\mu \bar U(p) W_+(x,p) U(p)}
\ee
where, henceforth, we will make explicit the distinction between the trace over 
the polarization states $\tr_2$ and the trace over the four spinorial indices $\tr_4$.
In the Weyl's representation, which is deeply connected to the theory of Lorentz'group 
represenations, these spinors can be written as \cite{Weinberg:1964cn,Moussa:1966gjd}:
\be\label{uvspin}
 U(p) = \sqrt{m} {D^S([p]) \choose D^S([p]^{\dagger -1})} \qquad 
 V(p) = \sqrt{m} {D^S([p]C^{-1}) \choose D^S([p]^{\dagger -1}C)}
\ee
with  $C=\ii \sigma_2$ ($\sigma_i$ being Pauli matrices).

The mean spin vector can now be calculated from \eqref{meansp4} by using the 
eq.~\eqref{spindensw} therein. We first observe that, since $S^\mu(p)$ is a real number,
we can also express it as: 
\begin{align}\label{spinreal}
  S^\mu(p) &= -\frac{1}{4m} \epsilon^{\mu\beta\gamma\delta} p_\delta 
  \left[ \tr_2 ( D^S([p]^{-1}) D^S( J_{\beta\gamma}) D^S([p]) \Theta(p)) +
   \tr_2 ( D^S([p]^{-1}) D^S( J_{\beta\gamma}) D^S([p]) \Theta(p))^* \right] \\
 \nonumber
& = -\frac{1}{4m} \epsilon^{\mu\beta\gamma\delta} p_\delta 
  \left[ \tr_2 ( D^S([p]^{-1}) D^S( J_{\beta\gamma}) D^S([p]) \Theta(p)) +
   \tr_2 ( D^S([p])^\dagger D^S( J_{\beta\gamma})^\dagger D^S([p]^{-1})^\dagger 
   \Theta(p)) \right] 
\end{align}
where we have taken advantage of the hermiticity of $\Theta(p)$ and the ciclicity
of the trace. We can now use the \eqref{spindensw2} and work out the numerator 
first:
\be\label{numtrac}
   \tr_2 ( D^S([p]^{-1}) D^S( J_{\beta\gamma}) D^S([p]) \bar U(p) W_+(x,p) U(p) +
   \tr_2 ( D^S([p])^\dagger D^S( J_{\beta\gamma})^\dagger D^S([p]^{-1})^\dagger 
   \bar U(p) W_+(x,p) U(p))  
\ee
where $U$ are the spinors defined in \eqref{uvspin}.
This expression can be written in a more compact and familiar form in the Dirac
spinorial formalism. We start by defining:
\be\label{Sigma}
  \Sigma_{\beta\gamma} = \left( \begin{array}{cc} D^S(J_{\beta\gamma}) 
  \; & \; 0 \\ 0 \; & \; D^S(J_{\beta\gamma})^\dagger \end{array} \right)
\ee
which is just the generator of Lorentz transformations written for the full spinorial 
representation $(0,1/2) \oplus (1/2,0)$ of the Dirac field, equal to
$(\ii/4) [\gamma_\beta,\gamma_\gamma]$. It can be readily seen that the \eqref{numtrac}
is equivalent to:
$$
   \frac{1}{m} \tr_2 ( \bar U(p) \Sigma_{\beta\gamma} U(p) \bar U(p) W_+(x,p) U(p)) 
$$
In general, if $A$ is a $2 \times 4$ and $B$ is a $4 \times 2$ matrix:
$$
  \tr_2 AB = \tr_4 BA
$$
hence the above trace can be rewritten:
$$
   \frac{1}{m} \tr_4 (\Sigma_{\beta\gamma} U(p) \bar U(p) W_+(x,p) U(p) \bar U(p))
$$
which can be worked out taking into account that:
$$
  U(p) \bar U(p) = \sum_r u_r(p) \bar u_r(p) = \slashed{p} + m
$$
Likewise, the denominator of \eqref{spindensw} can be rewritten as:
$$
 \tr_2 (\bar U(p) W_+(x,p) U(p)) = \tr_4 (W_+(x,p) U(p) \bar U(p)) = 
 \tr_4 ((\slashed{p} + m) W_+(x,p))
$$
Putting all together, we can write the mean spin vector as:
\be\label{meansp5}
  S^\mu(p) = -\frac{1}{4m^2} \epsilon^{\mu\beta\gamma\delta} p_\delta 
 \frac{\int \di \Sigma_\lambda p^\lambda \tr_4 (\Sigma_{\beta\gamma} (\slashed{p} + m) 
  W_+(x,p) (\slashed{p} + m))}
 {\int \di \Sigma_\lambda p^\lambda \tr_4 ((\slashed{p} + m) W_+(x,p))}
\ee

Likewise, one can also recast the equation \eqref{meansp4} - the exact expression 
of the mean spin vector at global equilibrium in the Boltzmann limit - for spin 1/2 
particles in the Dirac formalism. With the foregoing definitions and notations, the equation 
\eqref{spindensf2} can be rewritten as:
$$
  \Theta(p) = \frac{\bar U(p) \exp \left[ \frac{1}{2} \varpi: \Sigma \right] U(p)}
              {\tr_2 (\bar U(p) \exp \left[ \frac{1}{2} \varpi: \Sigma \right] U(p)) }
$$
which is a hermitian matrix. By using this equation, and the \eqref{spinreal}, 
\eqref{numtrac} the mean spin vector of \eqref{meansp4} can be rewritten by simply 
replacing the integrals of $W_+(x,p)$ with $\exp \left[ \frac{1}{2} \varpi: \Sigma \right]$ 
in the equation \eqref{meansp5}. We thus obtain:
\be\label{boltzexact1}
  S^\mu(p) =  -\frac{1}{4m^2} \epsilon^{\mu\beta\gamma\delta} p_\delta 
 \frac{ \tr_4 (\Sigma_{\beta\gamma} (\slashed{p} + m)\exp \left[ \frac{1}{2} \varpi: 
  \Sigma \right]  (\slashed{p} + m))} {\tr_4 ((\slashed{p} + m)
 \exp \left[ \frac{1}{2} \varpi: \Sigma \right])}
\ee
Taking into account that the trace of an odd number of gamma matrices vanishes and 
the commutator:
\be\label{commsigma}
 [\Sigma_{\beta\gamma},\gamma_\lambda] = - \ii g_{\beta\lambda} \gamma_\gamma 
 + \ii g_{\gamma\lambda} \gamma_\beta
\ee
it can be readily shown that the \eqref{boltzexact1} can be written in the simpler form:
\be\label{boltzexact2}
  S^\mu(p) =  -\frac{1}{2m} \epsilon^{\mu\beta\gamma\delta} p_\delta 
 \frac{ \tr_4 (\Sigma_{\beta\gamma} \exp \left[ \frac{1}{2} \varpi: 
  \Sigma \right])} {\tr_4 (\exp \left[ \frac{1}{2} \varpi: \Sigma \right])}
\ee
which looks certainly more suggestive and compact with respect to the general 
group-theoretical formula. 

Also the more general equation \eqref{meansp5} can be further developed and simplified. 
According to the \eqref{diracw}:
$$
 \slashed{p} \int \di \Sigma_\lambda p^\lambda W_+(x,p) = 
 m \int \di \Sigma_\lambda p^\lambda W_+(x,p)
$$
the \eqref{meansp5} becomes:
\be\label{meanspf}
  S^\mu(p) = -\frac{1}{2m} \epsilon^{\mu\beta\gamma\delta} p_\delta 
 \frac{\int \di \Sigma_\lambda p^\lambda \tr_4 (\Sigma_{\beta\gamma} W_+(x,p))}
 {\int \di \Sigma_\lambda p^\lambda \tr_4 W_+(x,p)}
\ee
which is already quite a suggestive formula. Furthermore, because of the \eqref{wignerint} 
we can write:
\begin{align*}
& \int \di \Sigma_\lambda p^\lambda  \tr_4 (\Sigma_{\beta\gamma} W_+(x,p)) = 
 \sum_{r,s} \frac{1}{4\varepsilon_p} 
 \delta(p^0 - \varepsilon_p) \media{\wadf{s}{p} \waf{r}{p}} \tr_4 (
 u_r(p) \bar u_s(p) \Sigma_{\beta\gamma}) \nonumber \\
& = 
 \sum_{r,s} \frac{1}{4\varepsilon_p} 
 \delta(p^0 - \varepsilon_p) \media{\wadf{s}{p} \waf{r}{p}}
 \bar u_s(p) \Sigma_{\beta\gamma} u_r(p)
\end{align*}
Now, by using the spinorial relation:
$$
 m \, \bar u_r(p) \Sigma^{\mu\nu} \gamma_\lambda u_s(p) = 
 \bar u_r(p) \Sigma^{\mu\nu} u_s(p) p_\lambda - 2 \ii \bar u_r(p) 
 (\gamma^\mu p^\nu - \gamma^\nu p^\mu) \gamma_\lambda u_s(p)
$$
the previous equation turns into:
\begin{align}\label{trint}
 \int \di \Sigma^\lambda p_\lambda  \tr_4 (\Sigma_{\beta\gamma} W_+(x,p))
 & = \sum_{r,s} \frac{1}{4\varepsilon_p^2} 
 \delta(p^0 - \varepsilon_p) \media{\wadf{s}{p} \waf{r}{p}}
  m \, \bar u_s(p) \Sigma_{\beta\gamma} \gamma_0 u_r(p) \nonumber \\
 & + 2 \ii \sum_{r,s} \frac{1}{4\varepsilon_p^2} 
 \delta(p^0 - \varepsilon_p) \media{\wadf{s}{p} \waf{r}{p}}
 \bar u_s(p) (\gamma_\beta p_\gamma - \gamma_\gamma p_\beta) \gamma_0
  \ u_r(p) 
\end{align}
The second term in the right hand side will not contribute to the mean spin vector
\eqref{meanspf} because of two momenta multiplying the Levi-Civita tensor. By using
the \eqref{commsigma}, the first term in the right hand side of \eqref{trint} can 
be rewritten:
\begin{align}\label{trint2}
  & \sum_{r,s} \frac{m}{4\varepsilon_p^2} \delta(p^0 - \varepsilon_p) \media{\wadf{s}{p} \waf{r}{p}}
  \, \left[ \bar u_s(p) \frac{1}{2} \{ \gamma_0, \Sigma_{\beta\gamma} \} 
  u_r(p) + \ii g_{0\gamma} \bar u_s(p) \gamma_\beta u_r(p) - 
  \ii g_{0 \beta} \bar u_s(p) \gamma_\gamma u_r(p) \right] \nonumber \\
 & = \sum_{r,s} \frac{m}{4\varepsilon_p^2} 
 \delta(p^0 - \varepsilon_p) \media{\wadf{s}{p} \waf{r}{p}}
  \left[ \bar u_s(p) \frac{1}{2} \{ \gamma_0, \Sigma_{\beta\gamma} \} 
  u_r(p) + 2 \ii \delta_{rs} ( p_\beta g_{0\gamma} - p_\gamma g_{0\beta})  \right] 
\end{align}
where we have used the relation
$$
   \bar u_s(p) \gamma^\lambda u_r(p) = 2 p^\lambda \delta_{rs}
$$
Again, the second term in the right hand side of \eqref{trint2} does not contribute 
to the mean spin vector because of the Levi-Civita tensor in \eqref{meanspf}. 
Now, since:
\be\label{axial}
 \{ \gamma_\lambda, \Sigma_{\beta\gamma} \} = \epsilon_{\sigma\lambda\beta\gamma} 
 \gamma^\sigma \gamma^5
\ee
we can finally rewrite the numerator of the mean spin vector \eqref{meanspf} as: 
\begin{align*}
& -\frac{1}{4} \epsilon^{\mu\beta\gamma\delta} \epsilon_{\sigma 0 \beta\gamma} 
  \, p_\delta \sum_{r,s} \frac{1}{4\varepsilon_p^2} \delta(p^0 - \varepsilon_p) 
 \media{\wadf{s}{p} \waf{r}{p}} \, \bar u_s(p) \gamma^\sigma \gamma^5 u_r(p) \\
& = p_0 \sum_{r,s} \frac{1}{8\varepsilon_p^2} \delta(p^0 - \varepsilon_p) 
 \media{\wadf{s}{p} \waf{r}{p}} \, \bar u_s(p) \gamma^\mu \gamma^5 u_r(p)
 - \delta^\mu_0 p_\sigma \sum_{r,s} \frac{1}{8\varepsilon_p^2} \delta(p^0 - \varepsilon_p) 
 \media{\wadf{s}{p} \waf{r}{p}} \, \bar u_s(p) \gamma^\sigma \gamma^5 u_r(p)
\end{align*} 
The second term vanishes because:
$$
  \bar u_s(p) \slashed p \gamma^5 u_r(p) = m \bar u_s(p) \gamma^5 u_r(p) = 0
$$
so we have, for the numerator of the \eqref{meanspf}:
$$
 \sum_{r,s} \frac{1}{8\varepsilon_p} \delta(p^0 - \varepsilon_p) 
 \media{\wadf{s}{p} \waf{r}{p}} \, \bar u_s(p) \gamma^\mu \gamma^5 u_r(p)
$$
This expression can be rewritten in form of an integral over an arbitrary hypersurface
of a divergence-free integrand, and so we get, by using again \eqref{wignerint}:
\be\label{meanspf2}
 S^\mu(p) = \frac{1}{2} \frac{\int \di \Sigma \cdot p \; \tr_4 (\gamma^\mu \gamma^5 W_+(x,p))}
 {\int \di \Sigma \cdot p \; \tr_4 W_+(x,p)}
\ee
Hence, the mean spin vector is proportional to the integral of the axial vector 
component of the covariant Wigner function over some arbitrary 3D space-like hypersurface.
Note that $S(p)$ is actually orthogonal to the four-momentum $p$ because 
$\tr_4 (\slashed p \gamma^5 W_+) = 0$ \cite{Vasak:1987um} (this can be shown also by using
the expansion \eqref{wignerdop}). This expression is consistent and extends the relation 
between $W(x,p)$ and $S^\mu(p)$ at $\cal O(\hbar)$ used in 
refs.~\cite{Fang:2016vpj,Weickgenannt:2019dks,Liu:2020flb} to determine the mean spin vector.

Finally, by using the inverse of the \eqref{axial}, that is:
$$
  \gamma^\mu \gamma^5 \delta_\lambda^\delta = \gamma^\delta \gamma^5 \delta^\mu_\lambda
  - \frac{1}{2} \epsilon^{\mu\delta\alpha\beta} \{ \gamma_\lambda, \Sigma_{\alpha\beta} \}
$$
and taking into account that $\tr_4 (\slashed p \gamma^5 W_+) = 0$
we obtain another form of the \eqref{meanspf2}:
\be\label{meanspf3}
  S^\mu(p) = -\frac{1}{4} \epsilon^{\mu\beta\gamma\delta} p_\delta 
 \frac{\int \di \Sigma_\lambda \tr_4 (\{ \gamma^\lambda, \Sigma_{\beta\gamma}\} 
 W_+(x,p))}{\int \di \Sigma_\lambda p^\lambda \tr_4 W_+(x,p)}
\ee
In the numerator, the educated reader shall recognize the matrix definining the 
canonical spin tensor of the Dirac field. However, it should be pointed out that 
the appearance of this combination does not imply the need of a particular spin tensor
to find the expression \eqref{meanspf3}, unlike originally stated in ref.~\cite{Becattini:2013fla}.
The point is that the polarization expression was obtained without any reference
whatsoever to the the tensors that are used to express the energy-momentum and 
angular momentum of the fields; this was already implied in the expressions of the
mean spin vector quoted in refs.~\cite{Florkowski:2017dyn,Florkowski:2018fap} and
will be discussed in more detail in section~\ref{polang}. 

Indeed, one could have chosen to express the relation between mean spin vector and 
covariant Wigner function through the \eqref{meanspf} or equally well with the 
eq.~\eqref{meanspf2}; they are completely equivalent forms of the \eqref{meanspf3}.

\section{Polarization from the angular momentum operator}
\label{polang}

We have seen how to calculate the mean spin vector from the spin density matrix 
definition in quantum field theory, see eq.~\eqref{spindens}. It is possible to 
calculate the same quantity with a different method. Assume that the total angular 
momentum tensor $J^{\mu\nu}$ 
can be decomposed on-shell in momentum space, so that we know the total angular momentum 
tensor of particles with given four-momentum $p$, say $\tilde J^{\mu\nu}(p)$. We could 
then be able to obtain the mean spin vector $S^\mu(p)$ by simply dividing this quantity 
by the number of particles with momentum $p$ and multiplying by the usual 
Levi-Civita like in the definition \eqref{luba}:
\be\label{spinalter}
  S^\mu(p) = -\frac{1}{2m} \epsilon^{\mu\nu\rho\sigma} p_\sigma 
   \frac{\tilde J_{\nu\rho}(p)}{\frac{\di N}{\di^3 p}}
\ee
This definition makes perfect sense as all particles with given momentum $p$ have 
the same rest frame and was indeed used in ref.~\cite{Becattini:2013fla} to derive
the expression of the mean spin vector at local thermodynamic equilibrium. We will
show that the \eqref{spinalter} leads to the eq.~\eqref{meanspf2} as well. 

The first step is to prove that $\tilde J_{\nu\rho}(p)$ exists and to find its form 
for the free Dirac field. Indeed, it is possible to show, under quite general assumptions, 
that conserved {\em charges} of free fields can be written as integrals over on-shell 
momentum four-vectors. In general, conserved charges can be written as integrals 
over a space-like 3D-hypersurface $\Sigma$ of a divergenceless current, also at 
operator level:
$$
   \wQ^{\mu_1 \ldots \mu_N} = 
\int_\Sigma \di \Sigma_\lambda \; \wJ^{\lambda \mu_1 \ldots \mu_N}
$$
With a suitable choice of the 3D boundaries, the above integral is independent of 
the hypersurface $\Sigma$ and it can then be calculated using any space-like $\Sigma$, e.g.  
$t=0$. Suppose now that the normal-ordered current $:\wJ:$ (normal ordering is 
necessary to have vanishing currents in vacuum) can be expressed as an integral 
over $k$ of some tensor functional of the covariant Wigner function $\wW(x,k)$:
\be\label{currentdec}
  :\wJ^{\lambda \mu_1 \ldots \mu_N}: \; = \int \di^4 k \; 
  {\cal F}[ \wW(x,k)]^{\lambda \mu_1 \ldots \mu_N} \;  
\ee
This is the case, for instance, for the charge current $:\wj^\mu:$ of the scalar field, for
which the functional would simply be $k^\mu \wW(x,k)$, or the current of the Dirac 
field for which it would be $\tr_4 (\gamma^\mu \wW(x,k))$, but it also applies to
stress-energy tensor and spin tensor, as we will see. Using eq.~(\ref{currentdec}), 
and taking advantage of the independence of the integration hypersurface, the 
charge can be written as: 
\begin{align}\label{construct}
  :\wQ^{\mu_1 \ldots \mu_N}: & = \int_\Sigma \di \Sigma_\lambda \;
 :\wJ^{\lambda \mu_1 \ldots \mu_N}: \;
  = \int_\Sigma \di \Sigma_\lambda \int \di^4 k \; 
 {\cal F}[\wW(x,k)]^{\lambda \mu_1 \ldots \mu_N} \\ \nonumber
 & = \int_{t=0} \di^3 \x \int \di^4 k \; {\cal F}[\wW(x,k)]^{0 \mu_1 \ldots \mu_N}
 = \int \di^4 k \; \left( \int_{t=0} \di^3 \x \; {\cal F}[\wW(x,k)]^{0 \mu_1 \ldots \mu_N} \right)
\end{align}
For {\em free fields}, the integration over $\x$ generally implies that the four-momentum 
$k$ is on shell. In fact, this depends on the specific form of the functional ${\cal F}$, 
but it holds for all cases of interest, and the proof is the same which led to the
equation~\eqref{wignerint}; after the last integration in $\di^3 \x$ a factor $\delta(k^2 - m^2)$ 
comes in which makes it possible to separate particle and antiparticle contribution 
and to express the generally conserved normal-ordered charge as:
\begin{align}\label{chargep}
 :\wQ^{\mu_1 \ldots \mu_N}: &= \int \di^4 k \; \delta (k^2-m^2) \wQ(k)^{\mu_1 \ldots \mu_N} 
 = \int \di^3 {\rm k} \left( \int \di k^0 \; \delta (k^2-m^2) \wQ (k)^{\mu_1 \ldots \mu_N} \right)
 \nonumber \\
 & \equiv \int \frac{\di^3 \kv}{2 \varepsilon_k} \; \left( \widehat q_+(k)^{\mu_1 \ldots \mu_N} 
 + \widehat q_-(k)^{\mu_1 \ldots \mu_N}  \right)
\end{align}
Altogether, the charge can be written as a sum over three-momenta of on-shell particles 
and antiparticles and a spectral decomposition in momentum space is obtained: 
\be\label{qdef}
   \frac{\di :\wQ^{\mu_1 \ldots \mu_N}:}{\di^3 \kv} 
   = \frac{1}{2\varepsilon_k} \widehat q_+(k)^{\mu_1 \ldots \mu_N} = 
   \int \di k^0 \int \di \Sigma_\lambda \; {\cal F}[\wW_+(x,k)]^{\lambda \mu_1 \ldots \mu_N} 
\ee
and likewise for antiparticles. The operators $\widehat q_\pm(k)$ are invariant by 
the addition of a total divergence to the current. For instance, for the vector current:
$$
  \wj^\lambda \to \wj^\lambda + \partial_\alpha \wA^{\lambda\alpha}
$$
where $\wA^{\lambda\alpha}$ is an anti-symmetric tensor, the corresponding 
$\widehat q_\pm$ get changed by:
$$
 \widehat q_\pm \to \widehat q_\pm + \int \di k^0 \int \di \Sigma_\lambda
 \partial_\alpha {\cal A}[\wW_\pm(x,k)]^{\lambda\alpha}
$$
where $\cal A$ is the suitable functional of the Wigner operator associated to
$\wA^{\lambda\alpha}$. The integral over the 3D hypersurface, for fixed $k$, can 
be turned into a boundary surface integral by means of the Stokes theorem and so,
provided that the suitable boundary conditions are enforced, vanishes.  

We now look for the spectral decomposition of the angular momentum-boost operators, 
hence the $\tilde J^{\mu\nu}(p)$ of the equation \eqref{spinalter}. The angular 
momentum-boost operator is the generator of the Lorentz transformations and can 
be written as:
\be\label{angmom}
  \wJ^{\mu\nu} = \int_\Sigma \di \Sigma_\lambda \; 
  \widehat{\cal J}^{\lambda,\mu\nu} = \int_\Sigma \di \Sigma_\lambda \; \left(
 x^\mu \wT^{\lambda\nu} - x^\nu \wT^{\lambda\mu} + \wspt^{\lambda,\mu\nu} \right)
\ee
with $\Sigma$ space-like hypersurface. There are two contributing terms: the so-called {\em orbital} 
part, depending on the stress-energy tensor, and the {\em spin tensor} operator $\wspt$. 
The generator in \eqref{angmom} is invariant under a so-called {\em pseudo-gauge 
transformation} of the stress-energy and spin tensor \cite{Hehl:1976vr} which amounts 
to add a divergence to the angular momentum-boost current $\widehat{\cal J}^{\lambda,\mu\nu}$. 
The choice of a stress-energy and a spin tensor is just a matter of convenience, and, for the
Dirac field, a convenient choice is the so-called {\em canonical} stress-energy and 
spin tensor:
\begin{align}\label{dcset}
 :\wT^{\mu\nu}(x): &= \frac{\ii}{2} : \Psibar(x) \gamma^\mu \codevnu \Psi(x): = 
 \int \di^4 k \; k^\nu \tr_4 (\gamma^\mu \wW(x,k)) \\ \nonumber
 :\wspt^{\lambda,\mu\nu}(x) : &= 
 \frac{1}{2} : \Psibar(x) \{ \gamma^\lambda, \Sigma^{\mu\nu} \} \Psi(x): 
 = \frac{1}{2} \int \di^4 k \; \tr_4 (\{ \gamma^\lambda, \Sigma^{\mu\nu} \} \wW(x,k))
\end{align}
where their relations with the covariant Wigner function have been written down.
Let us start with the {\em orbital} part of the angular momentum-boost operator:
\be\label{orbmom}
  :\widehat L^{\mu\nu}: \; \equiv \int \di\Sigma_\lambda \; \left( x^\mu :\wT^{\lambda\nu}:
 \,  -  x^\nu :\wT^{\lambda\mu}: \right)
 = \int \di\Sigma_\lambda \int \di^4 k \; \left( x^\mu k^\nu \tr_4 (\gamma^\lambda 
  \wW(x,k)) - x^\nu k^\mu \tr_4 (\gamma^\lambda \wW(x,k)) \right)
\ee
The subtlety here is that the functional ${\cal F}$, that we can write as:
$$
 {\cal F} = x^\mu k^\nu \tr_4 (\gamma^\lambda W(x,k)) - (\mu \leftrightarrow \nu)
$$
explicitely depends on $x$ and so the proof of the on-shellness of $k$ must be 
reviewed, what is done in detail in Appendix A. The result of this analysis is that
the orbital part of the angular momentum operator, with the canonical stress-energy
tensor in \eqref{dcset}, can be written as:
$$
  :\widehat L^{\mu\nu}: \; = \int \frac{\di^3 \kv}{2 \varepsilon_k} \; 
  \left( k^\mu \widehat G_+^\nu(k) - k^\nu \widehat G_+^\mu(k) \right) 
 + \;\; {\rm antiparticle\; term} 
$$
with $k$ on-shell and with $\widehat G$ a vector operator (see Appendix A). Thus,
the orbital part of the angular momentum does not contribute to the mean spin 
vector because of the Levi-Civita tensor which makes the orbital part vanishing.

On the other hand, the canonical spin tensor term in \eqref{angmom} has an algebraic 
dependence on the Wigner function, according to the \eqref{dcset} and the equation 
\eqref{qdef} can be applied with
$$
 {\cal F}[\wW_+(x,k)] = \frac{1}{2} \tr_4 \left( \{ \gamma^\lambda, \Sigma^{\mu\nu} \}
  \wW_+(x,k) \right)
$$
so the spin part of the total angular momentum-boost tensor is:
\be\label{spinmom}
  :\widehat S^{\mu\nu}: \;  = \int \di^4 k \int \di \Sigma_\lambda \;
  \tr_4 \left( \frac{1}{2} \{ \gamma^\lambda, \Sigma^{\mu\nu} \}  \wW_+(x,k) \right)
 + \;\; {\rm antiparticle\; term} 
\ee
Therefore, its contribution to the function $\tilde J^{\mu\nu}(p)$ for particles in 
\eqref{spinalter} is (with $k$ renamed $p$):
$$
  \tilde S_+^{\mu\nu}(p) = \int \di p^0 \int \di \Sigma_\lambda \;
  \tr_4 \left( \frac{1}{2} \{ \gamma^\lambda, \Sigma^{\mu\nu} \} W_+(x,p) \right)
$$
Now, in the equation \eqref{spinalter} we can replace $\tilde J^{\mu\nu}$ with the 
above expression and use the formula \eqref{momdens2} for the particle density in 
momentum space, obtaining:
\be\label{spinalter2}
  S^\mu(p) = -\frac{1}{4m} \epsilon^{\mu\nu\rho\sigma} p_\sigma 
  \frac{\int \di p^0 \int \di \Sigma_\lambda \; \tr_4 \left( 
  \{ \gamma^\lambda, \Sigma^{\nu\rho} \} W_+(x,p) \right) }{\frac{1}{2 m \varepsilon}
  \tr_4 w_+(p)} 
  = -\frac{1}{4} \epsilon^{\mu\nu\rho\sigma} p_\sigma 
  \frac{\int \di p^0 \int \di \Sigma_\lambda \; \tr_4 \left( 
  \{ \gamma^\lambda, \Sigma^{\nu\rho} \} W_+(x,p) \right) }{ 
   \int \di p^0 \int \di \Sigma \cdot p \; \tr_4 W_+(x,p)}
\ee
where we have used the eq.~\eqref{reducedwig} integrated in $p^0$. We can also 
recast the above formula by taking advantage of the cancellation of $\delta(k^0-\varepsilon_k)$ 
in the ratio:
$$
  S^\mu(p) = -\frac{1}{4} \epsilon^{\mu\nu\rho\sigma} p_\sigma 
  \frac{\int \di \Sigma_\lambda \; \tr_4 \left( 
  \{ \gamma^\lambda, \Sigma^{\nu\rho} \}  W_+(x,p) \right)}
   {\int \di \Sigma_\lambda p^\lambda \; \tr_4 W_+(x,p)}
$$
which is precisely the \eqref{meanspf2}. Hence, the method described in this section
leads to the same result obtained in section~\ref{polwig}.  

It is worth stressing the independence of the expression of $S^\mu(p)$ in 
eq.~\eqref{spinalter2} of the particular couple of stress-energy and spin tensor
chosen to calculate the total angular momentum spectral decomposition $\tilde J^{\mu\nu}(p)$. 
If we had used the Belinfante symmetrized tensor:
\be\label{belinf}
 :\wT^{\mu\nu}_B(x): = \frac{\ii}{4} : \Psibar(x) \gamma^\mu \codevnu \Psi(x)
+\Psibar(x) \gamma^\nu \codevmu \Psi(x): = \frac{1}{2} \int \di^4 k \; k^\nu 
  \tr_4 (\gamma^\mu \wW(x,k)) + k^\mu \tr_4 (\gamma^\nu \wW(x,k)) 
\ee
with associated vanishing spin tensor, for the derivation of the mean spin vector,
we would have obtained the same expression \eqref{spinalter2}. This happens because
the Belinfante associated ``orbital" angular momentum (which is actually the only 
term as $\wspt_B=0$) implies more terms in the decomposition with respect to the
equation \eqref{orbmom} (this is discussed at the end of Appendix A).

To conclude, as it was already discussed at the end of section~\ref{polwig}, the 
{\em expression} of the particle polarization as a function of momentum is independent of 
the pseudo-gauge transformation of stress-energy and spin tensor. For the Dirac 
field, the canonical stress-energy and spin tensor are actually the most convenient 
to obtain it by the method presented in this section, and yet, the same expression 
could be derived by using the Belinfante pseudo-gauge. The appearance of the canonical
spin tensor in the eq.~\eqref{meanspf3} does not give it a special physical meaning
and, indeed, the equivalenet forms \eqref{meanspf2},\eqref{meanspf} do not feature 
the canonical spin tensor. However, the {\em value} of the mean spin vector, as
well as any other quantity, may depend on the spin tensor because the density 
operator at local thermodynamic equiilibrium is sensitive to the pseudo-gauge 
transformations \cite{Becattini:2018duy,Becattini:2020riu}. Particularly, it is 
the Wigner function itself which acquires a dependence on the pseudo-gauge transformations 
through the density operator.

\section{Local thermodynamic equilibrium}
\label{wigloc}

We have seen in the previous sections how the spin density matrix and the mean spin 
vector relate to the covariant Wigner function. In turn, the covariant Wigner function 
depends on the density operator $\wrho$, see e.g. eq.~\eqref{wscalar2} and it is thus 
necessary to know the density operator to calculate it.

For a relativistic fluid which, at some time, is believed to have achieved local 
thermodynamic equilibrium, a powerful approach is the Zubarev's method of the stationary 
Non-Equilibrium Density Operator (NEDO) \cite{Zubarev:1979}. We refer the reader to the recent 
paper \cite{Becattini:2019dxo} for a detailed description. 

\begin{center}
\begin{figure}[ht]
\includegraphics[width=0.6\textwidth]{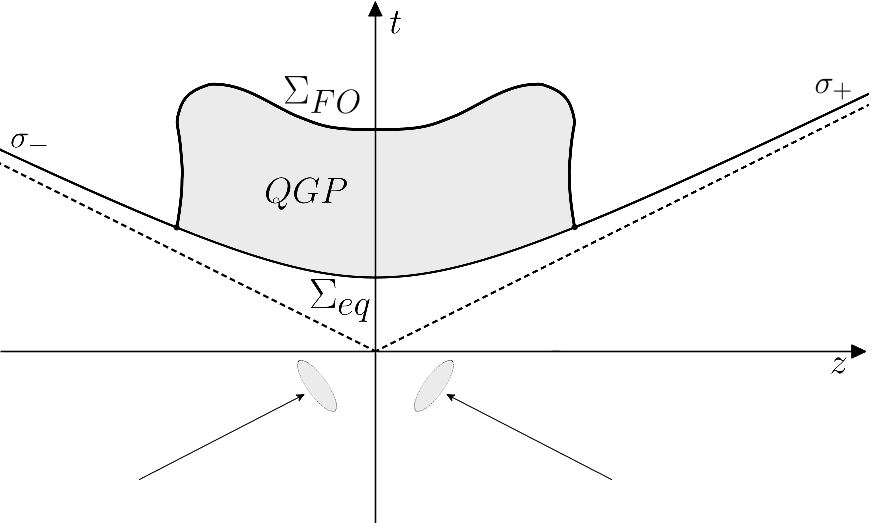}
\caption{Space-time diagram of a relativistic nuclear collision at very high 
energy. The hypersurface $\Sigma_{eq}$ corresponds to the achievement of local
thermodynamic equilibrium while $\Sigma_{FO}$ is the hypersurface where the Quark
Gluon Plasma decouples. The $\sigma_\pm$ are described in the text.} 
\label{fo}
\end{figure}
\end{center}

This approach is especially well suited for the physics of relativistic nuclear 
collisions, where the system supposedly achieves Local Thermodynamic Equilibrium (LTE) 
at some finite early ``time" (in the most used model, a finite hyperbolic time $\tau= 
\sqrt{t^2-z^2}$, see ref. \cite{Florkowski:2010zz}), to form a Quark Gluon Plasma
(QGP) which lives in a finite space-time region before breaking up at some 3D 
hypersurface $\Sigma_{FO}$ (see figure~\ref{fo}). 
The actual density operator, in the Heisenberg representation, must be a fixed, 
time and space independent, operator and for a fluid at local thermodynamic 
equilibrium it is obtained by maximizing the entropy $S=-\tr(\wrho \log \wrho)$
with the constraints of energy-momentum and charge densities \cite{Becattini:2014yxa}. 
The result is:
\be\label{densop}
  \wrho = \dfrac{1}{Z} 
  \exp \left[ -\int_{\Sigma_0} \di \Sigma \; n_\mu \left( \wT^{\mu\nu}(x) 
  \beta_\nu(x) - \zeta(x) \wj^\mu(x) \right) \right]
\ee
where $\beta$ is the four-temperature vector, $\zeta$ the ratio between chemical 
potential and temperature and $\Sigma_0$ is the initial 3D hypersurface where LTE 
is achieved. For relativistic nuclear collisions, this is supposedly the 3D hyperbolic 
hypersurface $\tau = \tau_0$, the $\Sigma_{eq}$ in figure~\ref{fo}. 
It should be pointed out that the form of the local equilibrium density operator
is pseudo-gauge dependent \cite{Becattini:2018duy}; the above form applies to the Belinfante 
stress-energy tensor only, so in the rest of the section it will be understood that
$\wT$ is the Belinfante symmetrized stress-energy tensor. 

However, the operator \eqref{densop}, as it stands, cannot be used to calculate 
the polarization of final state particles in practice. The reason is that the operators 
in the exponent of \eqref{densop} are to be evaluated at the time $\tau_0$, when 
the system is in the QGP phase and the field operators are those of the fundamental 
QCD degrees of freedom, quarks and gluons, whereas the creation and destruction 
operators in a formula such as \eqref{spindens} or \eqref{wignerdop} are clearly 
those of the hadronic asymptotic states, which can be expressed in terms of the 
effective hadronic fields. Even if we were able to write the effective hadronic 
fields in terms of the fundamental quark and gluon fields, those should be evaluated 
at different times, that is the initial ``time" $\tau_0$ and the decoupling time, 
so that the full dynamical problem of interacting quantum field should be solved. 
It is indeed convenient to rewrite $\wrhol(\tau_0)$ in terms of the operators at 
some present ``time" $\tau$ by means of the Gauss' theorem, taking into account that 
$\wT$ and $\wj$ are conserved currents \cite{Becattini:2019dxo}. Being:
$$
  \di \Sigma_\mu = \di \Sigma \, n_\mu
$$
where $\hat n$ is the unit vector perpendicular to the hypersurface and $\di \Omega$ 
being the measure of a 4D region in spacetime, we have
\be\label{gauss}
- \int_{\Sigma(\tau_0)} \!\!\!\!\!\! \di \Sigma_\mu \; \left( \wT^{\mu\nu} 
  \beta_\nu - \wj^\mu \zeta \right) = 
 - \int_{\Sigma(\tau)} \!\!\!\!\!\! \di \Sigma_\mu \; \left( \wT^{\mu\nu} 
  \beta_\nu - \wj^\mu \zeta \right) 
  + \int_\Omega \di \Omega \; \left( \wT^{\mu\nu} \nabla_\mu \beta_\nu - \wj^\mu 
 \nabla_\mu \zeta \right), 
\ee
where $\nabla$ is the covariant derivative. The region $\Omega$ is the portion of spacetime 
enclosed by the two hypersurface $\Sigma(\tau_0)$ and $\Sigma(\tau)$ and the timelike 
hypersurface at their boundaries, where the flux of ($\wT^{\mu\nu} \beta_\nu(x) - 
\wj^\mu \zeta(x)$) is supposed to vanish \cite{Becattini:2019dxo}. 
Consequently, the stationary NEDO reads:
\be\label{nedo}
 \wrho =  \dfrac{1}{Z} 
 \exp\left[ - \int_{\Sigma(\tau_0)} \!\!\!\!\!\! \di \Sigma_\mu \; \left( \wT^{\mu\nu} 
  \beta_\nu - \wj^\mu \zeta \right) \right] =
 \dfrac{1}{Z} 
 \exp\left[ - \int_{\Sigma(\tau)} \!\!\!\!\!\! \di \Sigma_\mu \; \left( \wT^{\mu\nu} 
  \beta_\nu - \wj^\mu \zeta \right) + \int_\Omega \di \Omega \; \left( \wT^{\mu\nu} 
  \nabla_\mu \beta_\nu - \wj^\mu \nabla_\mu \zeta \right) \right] 
\ee
In the case of heavy ion collisions the $\Sigma(\tau_0)$ - looking at figure~\ref{fo}
 - is the 3D hypersurface $\Sigma_{eq}$, while the hypersurface $\Sigma(\tau)$ is usually 
the joining of the freeze-out hypersurface $\Sigma_{FO}$ encompassing the QGP space-time 
region and the two side branches $\sigma_\pm$ subsets of the $\Sigma(\tau_0)$. A 
peculiarity of the heavy ion collisions is that the hypersurface of ``present" local 
equilibrium is partly time-like, that is $\hat n \cdot \hat n = -1$.
 
The density operator in \eqref{nedo} can be expanded perturbatively by identifying the 
two terms in its exponent: 
\be\label{aa}
 \wA = - \int_{\Sigma(\tau)} \!\!\!\!\!\! \di \Sigma_\mu \; 
  \left( \wT^{\mu\nu} \beta_\nu - \wj^\mu \zeta \right)
\ee
which is the supposedly, in hydrodynamics, the predominant term, and:
\be\label{bb}
  \wB = \int_{\Omega} \di \Omega \; \left( \wT^{\mu\nu} 
  \nabla_\mu \beta_\nu - \wj^\mu \nabla_\mu \zeta \right) 
\ee  
which is supposedly the small term. The $\wA$ and $\wB$ terms correspond to the 
LTE at the current time and the dissipative correction respectively. Hence, the 
leading term of the expansion of the mean value of any operator is the local equilibrium 
one, that is:
$$
 O \simeq \tr (\wrhol \widehat O) = \frac{\tr (\exp [\wA] \widehat O)}{\tr(\exp[\wA])}
$$
The convenient feature of this approach for the calculation of hydrodynamic constitutive 
equations, is the natural separation between non-dissipative terms - which are obtained 
by retaining the $\wA$ term - and the dissipative ones which are obtained by including 
$\wB$.

The calculation of $W(x,k)_{\rm LE}$, that is:
\be\label{wignerle1}
  W(x,k)_{\rm LE} = \frac{1}{Z_{\rm LE}}
   \Tr \left( \exp \left[ - \int_\Sigma \di \Sigma_\mu \; 
  \left( \wT^{\mu\nu}(y) \beta_\nu(y) - \wj^\mu(y) \zeta(y) \right) \right] \wW (x,k) 
  \right)
\ee
can be tackled by taking advantage of the supposedly slow variation of the fields 
$\beta$ and $\zeta$ in space-time compared with the variation of the Wigner operator
over microscopic scales. Beforehand, it should be
pointed out that the point $x$ where the Wigner function is to be evaluated is, to
a large extent, arbitrary. For, as we have seen in sections \ref{dirac} and \ref{polwig},
the 3D integration hypersurface of the Wigner function (see e.g. equation~\eqref{spindensw})
can be any hypersurface where the asymptotic hadronic fields are defined, one could 
choose a hyperplane at a sufficiently large value of the Minkowski time $t$ so as
to be completely outside the QGP spacetime region (see fig.~\ref{fo}), where hadronic 
fields cannot be used. However, this is not a convenient choice: at large times 
the fields $\beta$ and $\zeta$ are no longer defined because the system is not a
fluid anymore and it would then be difficult to estimate the Wigner function therein. 
A much better choice is an equivalent (from the viewpoint of the Gauss theorem) 
3D hypersurface encompassing the QGP and much closer to where the hydrodynamic fields 
are still defined. This hypersurface $\Sigma$ can be obtained by joining of the 
break-up hypersurface $\Sigma_{FO}$ and the two branches $\sigma_\pm$, as discussed above.  
Now one can evaluate $W(x,k)_{\rm LE}$ in space-time points where $\beta$ and $\zeta$
exist, with the exception of the branches $\sigma_\pm$ where the matter is not 
a fluid. Indeed, those branches involve the cold nuclear matter not participating
the QGP formation and its contribution is usually neglected. 
Since the hydrodynamic-thermodynamic fields $\beta$ and $\zeta$ are slowly varying, one can 
expand them in a Taylor series around $x$, the point where the Wigner operator is 
evaluated, and, retaining only the first order:
$$
 \beta_\nu(y) \simeq \beta_\nu(x) + \partial_\lambda \beta_\nu(x) (y-x)^\lambda
$$
and similarly for $\zeta$. Inserting in the eq.~\eqref{wignerle1}:
\begin{align}\label{wignerle2}
  W(x,k)_{\rm LE} & \\ \nonumber
 & \simeq \frac{1}{Z_{\rm LE}}
  \Tr \left( \exp \left[ - \int_\Sigma \di \Sigma_\mu \; 
  \left( \wT^{\mu\nu}(y) [ \beta_\nu(x) + \partial_\lambda\beta_\nu(x)(y-x)^\lambda ] 
  - \wj^\mu(y) [\zeta(x) +\partial_\lambda \zeta(x) (y-x)^\lambda] \right) \right] 
  \wW (x,k) \right) \\ \nonumber
 & = \frac{1}{Z_{\rm LE}} \Tr \left( \exp \left[ - \beta_\nu(x) \int_\Sigma 
  \di \Sigma_\mu \; \wT^{\mu\nu}(y) - \partial_\lambda \beta_\nu(x) 
 \int_\Sigma \di \Sigma_\mu \; (y-x)^\lambda \wT^{\mu\nu}(y) 
  - \zeta(x) \int_\Sigma  \di \Sigma_\mu \;\wj^\mu \right.\right. \\ \nonumber
 & \left.\left.  - \partial_\lambda \zeta(x) \int_\Sigma \di \Sigma_\mu \; 
  (y-x)^\lambda \wj^\mu \right] \wW (x,k) \right) 
\end{align}

This approximation corresponds to the hydrodynamic limit, where the mean value of 
local operators is determined by the local values of the thermodynamic fields. 
The gradient of $\beta$ in the last equation can be split into the symmetric and 
the anti-symmetric part giving rise to:
\be\label{symmanti}
 - \frac{1}{2} \varpi_{\lambda\nu} \int_\Sigma \di \Sigma_\mu \; 
  (y-x)^\lambda \wT^{\mu\nu}(y) - (y-x)^\nu \wT^{\mu\lambda}(y)
 + \frac{1}{4} (\partial_\lambda \beta_\nu + \partial_\nu \beta_\lambda) 
  \int_\Sigma \di \Sigma_\mu \; 
    (y-x)^\lambda \wT^{\mu\nu}(y) + (y-x)^\nu \wT^{\mu\lambda}(y)
\ee
where $\varpi$ is the thermal vorticity \eqref{thvort}.
We can recognize in the first term of the above equation the total angular momentum
operator, with a proviso: the above integration is over a 3D hypersurface 
$\Sigma \supset \Sigma_{FO}$ which is not fully space-like, in fact it has a time-like 
part. Notwithstanding, as the angular momentum-boost current is divergenceless and
being $\Sigma = (\Sigma_{FO} \cup \sigma_\pm)$ as discussed, we can again use the 
Gauss theorem and write:
$$
   \int_\Sigma \di \Sigma_\mu \; 
  (y-x)^\lambda \wT^{\mu\nu}(y) - (y-x)^\nu \wT^{\mu\lambda}(y)
  =  \int_{\Sigma_{eq}}\!\!\!\!\! \di \Sigma_\mu \; 
  (y-x)^\lambda \wT^{\mu\nu}(y) - (y-x)^\nu \wT^{\mu\lambda}(y)
$$ 
where $\Sigma_{eq}$ is the initial, space-like local thermodynamic equilibrium. The 
latter is, by definition, the conserved total angular momentum-boost generator with 
center $x$, that is $\wJ_x^{\mu\nu}$.

The main contribution to the Wigner function supposedly arises from the terms surviving at
the global equilibrium, occurring when $\partial_\mu \beta_\nu + \partial_\nu \beta_\mu =0$
and $\partial_\mu \zeta = 0$. However, the symmetric term in \eqref{symmanti} as 
well as the $\partial \zeta$ term in \eqref{wignerle2} may in principle contribute
at LTE (they are non-dissipative, non-equilibrium terms) and it would be interesting 
to assess their quantitative effect. Assuming that they are negligible, we have:
\be\label{wignerge}
 W(x,k)_{\rm LE} \simeq \frac{1}{Z} \Tr \left( \exp \left[ - \beta_\nu(x) 
 \wP^\nu + \frac{1}{2} \varpi_{\nu\lambda}(x) \wJ^{\nu\lambda}_x + \zeta(x) \wQ \right] 
 \wW (x,k) \right)
\ee
This expression is the {\em global thermodynamic equilibrium} mean of the Wigner 
function $W(x,k)_{\rm GE}$ with four-temperature and thermal vorticity values just
equal to their values in the point $x$ where the Wigner function is to be evaluated.

\subsection{Polarization at local thermodynamic equilibrium}
\label{polaloc}

Working out \eqref{wignerge} is, in principle, much easier than a complete local 
equilibrium calculation and yet, the exact form has not been determined so far. 
A possible approach is linear response theory, taking as the term 
$\varpi_{\lambda\nu}(x) \wJ_x^{\lambda\nu}$ in eq.~\eqref{wignerle2}
as the small term compared to the main term $ - \beta_\nu(x) \wP^\nu + \zeta(x) \wQ$.
This method, however, involves the calculation of complicated integral correlators 
between the angular momentum operator and the Wigner operator and, although viable, has 
never been attempted in literature. 

In ref.~\cite{Becattini:2013fla} an educated {\em ansatz} was introduced based 
on the on-shell De Groot's approximation of the general form of the covariant Wigner 
function:
\be\label{wigapprox}
 W(x,k) \simeq \frac{1}{2} \sum_{r,s} \int \frac{\di^3 \p}{\varepsilon} \; 
 \delta^4(k-p) u_r(p) f(x,p)_{rs} \bar u_s(p) - 
 \delta^4(k+p) v_r(p) \bar f(x,p)_{s r} \bar v_s(p)
\ee 
where $f_{rs}(x,p)$ is a $2 \times 2$ distribution function and $r$ $s$ label spin 
states. Now we know that the Boltzmann limit of \eqref{spindensw2} must yield the 
spin density matrix \eqref{spindensf}, i.e. by using the \eqref{uvspin},
\be\label{boltzlim}
 \frac{\int \di \Sigma_\mu p^\mu \bar U(p) W_+(x,k)_{\rm GE} U (p)}
 {\tr_2 \int \di \Sigma_\mu p^\mu \bar U(p) W_+(x,k)_{\rm GE} U(p)} \longrightarrow
  \frac{\bar U(p) \exp\left[ \frac{1}{2} \varpi: \Sigma \right] U(p)}{\tr_2(
  \bar U(p) \exp\left[ \frac{1}{2} \varpi: \Sigma \right] U(p) )}
\ee
Hence, a suitable form of $f$ was assumed giving the correct Boltzmann (as well 
as the non-relativistic) limit:
\be\label{pspace}
  f_{rs}(x,p) = \bar u_r(p) \exp \left[ \beta \cdot p - 
  \frac{1}{2} \varpi: \Sigma + I \right]^{-1} u_s(p)
\ee
The equations \eqref{wigapprox} and \eqref{pspace} together lead to the following 
form of the mean spin vector for spin 1/2 particles \cite{Becattini:2013fla}:
\be\label{basic}
 S^\mu(p)= - \frac{1}{8m} \epsilon^{\mu\rho\sigma\tau} p_\tau 
 \frac{\int_{\Sigma_{FO}} \di \Sigma_\lambda p^\lambda n_F (1 -n_F) \varpi_{\rho\sigma}}
  {\int_{\Sigma_{FO}} \di \Sigma_\lambda p^\lambda n_F}
\ee
where $n_F$ is the Fermi-Dirac phase-space distribution function:
$$
  n_F = \frac{1}{\exp[\beta\cdot p - \mu q]+1}
$$
$q$ being a charge of the particle and $\mu$ the corresponding chemical potential.
The \eqref{basic} is, in the Boltzmann limit, in full agreement with the equation 
\eqref{formula} which is the first order formula obtained within a single particle 
framework. The problem to determine the exact form at global equilibrium including 
quantum statistics effects is - as mentioned - yet to be solved.

\section{Summary and outlook}
\label{outlook}

The calculation of polarization in a relativistic fluid stands out as a fascinating 
endeavour in quantum field theory. As we have seen, it requires the use of a broad 
range of concepts and theoretical tools and it involves intriguing fundamental physics 
problems such as the physical significance of the spin tensor. It should be emphasized 
that it is not just an academic problem: polarization in the QCD plasma has been 
observed in experiments and much of its phenomenological potential as a probe of 
the hot QCD matter is still to be explored. In this regard, much theoretical 
and experimental work is ongoing. For a comprehensive review of the status of the
subject, we refer the reader to the recent review \cite{Becattini:2020ngo}.

The formula \eqref{basic} is the benchmark for most estimates of polarization. 
While very successful in reproducing the global polarization of $\Lambda$ hyperons
in relativistic heavy ion collisions, a disagreement with the data was found as to 
the momentum dependence of polarization \cite{Becattini:2020ngo}. These discrepancies could 
be an effect of incorrect hydrodynamic initial conditions, resulting in a distorted 
thermal vorticity field at the freeze-out or they could possibly arise from missing 
theoretical ingredients and major corrections to the equation \eqref{formula}, which 
is a leading order formula in thermal vorticity. Even though thermal vorticity is 
apparently a small number in relativistic heavy ion collisions, a quantitative role 
of the yet unknown exact formula of the Wigner function \eqref{wignerle2} cannot 
be ruled out for the present. Similarly, dissipative corrections to the mean spin 
vector are quantitatively unknown thus far. Even the estimate of the first order 
correction to the formula \eqref{basic} in the linear approximation with the operator 
\eqref{bb} is a formidable task as it involves, in heavy ion collisions, the full 
non-perturbative QCD regime (there is an ongoing effort in this direction 
\cite{Hattori:2019lfp}). The theory of the polarization in relativistic fluid is
still to be fully developed.

\section*{Acknowledgments}

Stimulating discussions with W. Florkowski and L. Tinti are gratefully 
acknowledged. I am very grateful to Q. Wang and X. G. Huang for very useful 
suggestions and to M. Buzzegoli and A. Palermo for a careful review 
of the manuscript and for making the figure. Special thanks to Enrico Speranza
for very valuable remarks and comments. 

\bibliography{./mybib}

\appendix

\section{Angular momentum decomposition}

We shall prove that the orbital part of the angular momentum operator \eqref{orbmom}
can be written as an integral in momentum space of on-shell functions. We will confine
ourselves to the proof for the particle term in \eqref{wignerdop}, its extension to
the anti-particle term and the proof of the vanishing of the mixed term being alike.
By using the \eqref{wignerdop} and choosing the hyperplane $t=0$ as integration 
hypersurface, for the particle term, we can write:
\begin{align}\label{App1}
 & \int \di^3 \x \; x^\mu k^\nu \tr_4 (\gamma^0 \wW(x,k)) \\
 & = \int \di^3 \x \; x^\mu k^\nu \sum_{r,s} 
 \frac{1}{(2\pi)^3} \int \frac{\di^3 \p}{2\varepsilon} \frac{\di^3 \p^\prime}
 {2\varepsilon^\prime} \e^{-\ii (p - p^\prime) \cdot x} 
 \delta^4 \left( k - \frac{p+p^\prime}{2} \right) 
  \wadf{s}{p^\prime} \waf{r}{p} \bar u_s(p') \gamma^0 u_r(p)
  \\
 & = 8 \int \di^3 \x \; x^\mu k^\nu \sum_{r,s} 
 \frac{1}{(2\pi)^3} \int \frac{\di^3 \p}{4\varepsilon \, \varepsilon_{k,p}} 
  \e^{\ii ( 2 k -2 p )\cdot x} \delta \left( k^0 - \frac{\varepsilon
  +\varepsilon^\prime_{k,p}}{2} \right) 
  \wadf{s}{2k-p} \waf{r}{p} \bar u_s(2k - p) \gamma^0 u_r(p)
\end{align}
where:
$$
  \varepsilon_{k,p} = \sqrt{(2 {\bf k} - {\bf p})^2 + m^2} 
$$
and it is understood that in the arguments of creation and destruction operators, as
well as of spinors $u$, only the spatial part of the four-vector $k$, that is
${\bf k}$, enters. 

For $\mu=0$, the integration is straightforward as $x^0$ is constant on the hyperplane
and we get, after integrating in $\di^3 \x$:
\begin{align*}
 & 8 x^0 k^\nu \sum_{r,s} \int \frac{\di^3 \p}
  {4\varepsilon \, \varepsilon_{k,p}} \delta^3 ({\bf p}-{\bf k}) 
  \delta \left( k^0 - \frac{\varepsilon +\varepsilon_{k,p}}{2} \right) 
  \wadf{s}{2k-p} \waf{r}{p} \bar u_s(2k - p) \gamma^0 u_r(p) \\
& = x^0 k^\nu \sum_{r,s} \frac{1}{(2\pi)^3} 
  \frac{1}{4\varepsilon_k^2} \delta (k^0 -\varepsilon) \wadf{s}{k} \waf{r}{k}
  \bar u_s(k) \gamma^0 u_r(k) = x^0 k^\nu \delta ( k^0 -\varepsilon_k) 
   \sum_{r} \frac{1}{2\varepsilon_k} \wadf{r}{k} \waf{r}{k}
\end{align*}
because $p'=2k-p$ and ${\bf p}={\bf k}$ implies in turn $k=p=p^\prime$, hence $k$ 
is on-shell; we have also used the known spinor relations.

For $\mu= i \ne 0$ we can replace $x^\mu$ with a derivative of the exponential 
and, integrating by parts:
\begin{align*}
 & \int \di^3 \x \; x^i k^\nu \tr_4 (\gamma^0 \wW(x,k)) \\
& =  4 \ii \int \di^3 \x \; k^\nu \sum_{r,s} 
 \frac{1}{(2\pi)^3} \int \frac{\di^3 \p}{4\varepsilon \, \varepsilon_{k,p}} 
  \frac{\partial}{\partial p^i} \e^{\ii (2 k -2 p) \cdot x} 
  \delta \left( k^0 - \frac{\varepsilon +\varepsilon_{k,p}}{2} \right) 
  \wadf{s}{2k-p} \waf{r}{p} \bar u_s(2k - p) \gamma^0 u_r(p)
\\
& = 4 \ii \int \di^3 \x \; k^\nu \sum_{r,s} 
 \frac{1}{(2\pi)^3} \int \frac{\di^3 \p}{4\varepsilon \, \varepsilon_{k,p}} 
 \frac{\partial}{\partial p^i} \left[ \e^{\ii (2 k - 2 p) \cdot x} 
  \delta \left( k^0 - \frac{\varepsilon +\varepsilon_{k,p}}{2} \right) 
  \wadf{s}{2k-p} \waf{r}{p} \bar u_s(2k - p) \gamma^0 u_r(p)
  \right]
 \\
& - 4 \ii \int \di^3 \x \; k^\nu \sum_{r,s} 
 \frac{1}{(2\pi)^3} \int \frac{\di^3 \p}{4\varepsilon \, \varepsilon_{k,p}} 
  \e^{\ii (2 k -2 p) \cdot x } \frac{\partial}{\partial p^i}
  \left[ \delta \left( k^0 - \frac{\varepsilon +\varepsilon_{k,p}}{2} \right) 
  \wadf{s}{2k-p} \waf{r}{p} \bar u_s(2k - p) \gamma^0 u_r(p) \right]
\end{align*}
The first term gives rise to a boundary integral which vanishes for fixed $k$ and
only the second term survives. We can now integrate in $\di^3 \x$ getting:
\be\label{intspace}
 - \frac{\ii}{2} k^\nu \sum_{r,s} 
  \int \frac{\di^3 \p}{4\varepsilon \varepsilon_{k,p}} 
  \delta^3({\bf p}-{\bf k}) \frac{\partial}{\partial p^i}
  \left[ \delta \left( k^0 - \frac{\varepsilon +\varepsilon_{k,p}}{2} \right) 
  \wadf{s}{2k-p} \waf{r}{p} \bar u_s(2k - p) \gamma^0 u_r(p) \right]
\ee
There appear two derivative terms in the above expression: the derivative of 
the $\delta$ can be written as:
\be\label{deriv1}
  \frac{\partial}{\partial p^i} \delta \left( k^0 - \frac{\varepsilon +
  \varepsilon^\prime_{k,p}}{2} \right) =
  - \frac{1}{2} \frac{\partial}{\partial k^0} \delta \left( k^0 - \frac{\varepsilon +
  \varepsilon_{k,p}}{2} \right) \frac{\partial}{\partial p^i} (\varepsilon+
  \varepsilon_{k,p}) = - \frac{1}{2} \frac{\partial}{\partial k^0} 
  \delta \left( k^0 - \frac{\varepsilon + \varepsilon_{k,p}}{2} \right)
  \left( \frac{p^i}{\varepsilon} - \frac{2 k^i - p^i}{\varepsilon_{k,p}} \right)
\ee
while the derivative of the factor including creation and destruction operators and
spinors yields, taking into account the $\delta^3({\bf p}-{\bf k})$:
\begin{align}\label{deriv2}
 & \delta^3({\bf p}-{\bf k}) \frac{\partial}{\partial p^i} \wadf{s}{2k-p} 
  \waf{r}{p} \bar u_s(2k - p) \gamma^0 u_r(p) \\ \nonumber
 & = \delta^3({\bf p}-{\bf k}) \left[ \left( \wadf{s}{p} \codevi \waf{r}{p} \right) 
  \bar u_s(p) \gamma^0 u_r(p)+  
  \wadf{s}{p}\waf{r}{p} \left( \bar u_s(p) \codevi \gamma^0 u_r(p) \right)
 \right]
\end{align}
We can now plug the equations \eqref{deriv1} and \eqref{deriv2} into the \eqref{intspace}. 
The term \eqref{deriv1} vanishes because:
\begin{align*}
 & \delta^3({\bf p}-{\bf k}) \left( \frac{p^i}{\varepsilon} - 
\frac{2 k^i - p^i}{\varepsilon_{k,p}} \right) k^\nu \frac{\partial}{\partial k^0} 
  \delta \left( k^0 - \frac{\varepsilon + \varepsilon_{k,p}}{2} \right)
 = - \delta^3({\bf p}-{\bf k})  \left( \frac{p^i}{\varepsilon} - \frac{2 k^i - 
   p^i}{\varepsilon_{k,p}} \right) \delta^\nu_0 
  \delta \left(k^0 - \frac{\varepsilon + \varepsilon_{k,p}}{2} \right)  \\
& = - \delta^3({\bf p}-{\bf k}) \delta^\nu_0 
  \delta \left(k^0 - \varepsilon \right)
  \left( \frac{k^i}{\varepsilon_k} - \frac{k^i}{\varepsilon_k} \right) = 0
\end{align*} 
and we are just left with the term from \eqref{deriv2}. 

We can now integrate in $\di^4 k$ according to the equation \eqref{orbmom}. For
$\mu=i$:
\begin{align}\label{App2}
&  \int \di^4 k \int \di^3 \x \; x^i k^\nu \tr_4 (\gamma^0 \wW(x,k)) \nonumber \\
& = - \frac{\ii}{2} \int \di^4 k \; k^\nu \sum_{r,s} \frac{1}{4\varepsilon_k^2} 
 \delta(k_0 - \varepsilon_k )  \left[ \left( \wadf{s}{k} \codevki \waf{r}{k} \right) 
  \bar u_s(k) \gamma^0 u_r(k) +  \wadf{s}{k}\waf{r}{k} \left( 
  \bar u_s(k) \codevki \gamma^0 u_r(k) \right) \right] \nonumber \\
& = - \frac{\ii}{2} \int \di^3 \kv \; k^\nu \sum_{r,s} \frac{1}{4\varepsilon_k^2} 
  \left[ \left( \wadf{s}{k} \codevki \waf{r}{k} \right) 
  \bar u_s(k) \gamma^0 u_r(k) +  \wadf{s}{k}\waf{r}{k} \left( 
  \bar u_s(k) \codevki \gamma^0 u_r(k) \right) \right] 
\end{align}
with $k^\nu$ again on-shell. We can then conclude that:
$$
 \int \di^4 k \int_{t=0} \di^3 \x \; x^\mu k^\nu \tr_4 (\gamma^0 \wW(x,k)) =
  \int \frac{\di^3 \kv}{2 \varepsilon_k} \; \widehat G^\mu (k) k^\nu
$$
with $k$ on-shell and $\widehat G^0(k) = 0$ if $x^0 = 0$ is chosen.

Finally, we briefly address the calculation of the angular momentum tensor by using
the Belinfante stress-energy tensor \eqref{belinf} where only the orbital part is
involved. The calculation is very similar to the one just described, with the important
difference that the second term of \eqref{belinf}, obtained by swapping the indices of 
the first term in \eqref{belinf}, leads to a term akin to the left hand side of the 
eq.~\eqref{App1} with exchanged indices:
$$
  \int \di^3 \x \; x^\mu k^0 \tr_4 (\gamma^\nu \wW(x,k))
$$
However, the final result is not proportional to $k^\nu$ and a double derivative 
term appears just like in eq.~\eqref{App2}; therefore, this term is not cancelled 
by the Levi-Civita tensor in the calculation of the mean spin, unlike in the 
canonical case.

\end{document}